\documentclass[fleqn,usenatbib]{mnras}

\usepackage{ulem}

\usepackage[T1]{fontenc}

\DeclareRobustCommand{\VAN}[3]{#2}
\let\VANthebibliography\thebibliography
\def\thebibliography{\DeclareRobustCommand{\VAN}[3]{##3}\VANthebibliography}

\usepackage{graphicx}	
\usepackage{amsmath}	
\usepackage{amssymb}	
\usepackage{siunitx}    
\usepackage{caption}    
\usepackage{subcaption} 

\newcommand\psj{PSJ}

\title[Aspect Change Effects in NEA Phase Curves]{The Effect of Aspect Changes on Near-Earth Asteroid Phase Curves}

\author[S. L. Jackson et al.]{
S. L. Jackson,$^{1}$\thanks{E-mail: samuel.jackson@open.ac.uk (SLJ)}
B. Rozitis,$^{1}$
L. R. Dover,$^{2}$
S. F. Green,$^{1}$
U. C. Kolb,$^{1}$
A. E. Andrews,$^{1}$
S. C. Lowry$^{2}$
\\
$^{1}$School of Physical Sciences, The Open University, Milton Keynes, MK7 6AA, UK\\
$^{2}$Centre for Astrophysics and Planetary Science, University of Kent, Canterbury, CT2 7NH, UK
}

\date{Accepted XXX. Received YYY; in original form ZZZ}

\pubyear{2022}

\usepackage{newtxtext,newtxmath}
\begin{document}
\label{firstpage}
\pagerange{\pageref{firstpage}--\pageref{lastpage}}
\maketitle

\begin{abstract}
    Phase curves of asteroids are typically considered to depend solely on the scattering properties of airless particulate surfaces and the size of the object being studied.
    In this study, we demonstrate the additional dependence of phase curves on object shape, rotation pole orientation, and viewing geometry over an apparition.
    Variations in the phase curve of near-Earth asteroid (159402) 1999 AP10 over its apparition from July 2020 - January 2021 are verified to be due to aspect changes over the apparition.
    This is achieved through shape modelling of the asteroid and simulation of the phase curve over the apparition.
    We present simulations of asteroid phase curves over a range of geometries to understand the potential magnitude of this aspect effect, and under which circumstances it can begin to dominate in the phase curves.
    This dependence on aspect may introduce significant additional uncertainty in the properties derived from phase curve data.
    We provide and demonstrate software code to estimate the aspect-related uncertainty in near-Earth asteroid phase curves through simulation and model fitting of a randomly generated sample of ellipsoidal asteroid models over the observed viewing geometry.
    We demonstrate how ignoring this effect may lead to misleading interpretations of the data and underestimation of uncertainties in further studies, such as those in the infrared that use phase curve derived parameters when fitting physical properties of an asteroid.
\end{abstract}




\section{Introduction}\label{sec:introduction}
    
    As the heliocentric distance, geocentric distance, and phase angle (defined as the angle between the Earth and Sun position vectors as viewed from the asteroid) of an asteroid change, the observed brightness of the asteroid viewed from Earth will change.
    To make meaningful comparisons of the brightness of different asteroids, it is standard practice to convert the apparent magnitude to that at a distance of $1$\,au from both the Earth and the Sun, referred to as the reduced magnitude.
    The variation of the reduced magnitude of an asteroid (averaged over any rotational light curve period) with changing phase angle is known as the phase curve.
    Phase curves allow us to predict the apparent magnitude of an asteroid for observation planning, and gain insight into the physical properties of these objects.
    The absolute magnitude ($H$) of an asteroid, defined as the mean reduced magnitude when observed at a phase angle of $0\si\degree$, is related to the diameter and geometric albedo of the object.
    The slopes of phase curves are typically explained solely by the scattering properties of the material on asteroid surfaces.
    Models such as the $H$, $G$ \citep{1989aste.book...524}, $H$, $G_1$, $G_2$ \citep{2010Icar..209..542M}, and $H$, $G_{12}$ \citep{2010Icar..209..542M} systems aim to approximate the behaviour of these scattering properties in a parametric manner.
	Fitting these models to photometric data helps us to constrain the absolute magnitudes, geometric albedos, and potential taxonomic classifications of asteroids \citep{1989aste.book...524,2012Icar..219..283O,2016P&SS..123..117P}.
	
    These parametric models are unable to account for changes in brightness of asteroids due to changing viewing aspect either during or between apparitions.
    Changing viewing geometry changes the cross-sectional area visible to the observer for an irregularly shaped asteroid.
    This can impart brightness variations onto asteroid phase curves \citep{rozitis2020}.
    This means that a single set of model parameters may not accurately represent the phase curve in other apparitions.
    Aspect effects are likely to be significant for asteroids that are irregularly shaped and experience large changes to viewing aspect during their observable apparitions, e.g. near-Earth asteroids (NEAs).
    Shape-induced modulations to phase curves may limit the taxonomic information that can be extracted from fits using parametric phase curve models.
    The usage of the $G_{12}$ parameter alone to estimate the classifications of individual asteroids has been shown to provide poor results \citep{2012Icar..219..283O}, and aspect effects would compound these issues further.
    
    Sparse observations of near-Earth asteroids used to construct phase curves (for example from large scale surveys) that are affected by aspect effects will result in significant errors introduced into estimates of absolute magnitudes and slope parameters ($G$, $G_1$ and $G_2$, or $G_{12}$).
    This may cause large errors to propagate into diameter estimates and predictions of future brightness.
    Errors in phase curve properties of asteroids hinder accurate derivation of thermophysical properties of their surfaces through models such as the Near-Earth Asteroid Thermophysical Model \citep[NEATM;][]{1998Icar..131..291H} or the Advanced Thermophysical Model \citep[ATPM;][]{2011MNRAS.415.2042R}.
    The presence of large-scale errors in sizes of individual objects also has natural consequences for planetary defence mitigation planning if only photometric data are available (i.e. the object is initially too distant for planetary radar).
    
    The definitions of ``reference phase curves'' \citep{2001Icar..153...37K} and ``proper phase curves'' \citep{2020A&A...642A.138M} make use of convex shape models and rotation pole solutions to correct phase curves to equatorial viewing geometry, following on from similar work using triaxial ellipsoids by \citet{1988Icar...76...19D}.
    \citet{2020PSJ.....1...73M} extract the ``reference phase curve'' for the Lucy mission target (11351) Leucus in order to remove shape and viewing geometry effects from the results using a derived shape model from photometric light curve and occultation observations.
    Inversion of a large collection of ground-based and Gaia asteroid photometric data was performed by \citet{2021A&A...649A..98M} to correct for aspect changes to the phase curves through: extraction of the intrinsic shape and scattering properties of the surface, simulation of the ``proper phase curve'', and subsequent fitting of the $H$, $G_1$, $G_2$ system to the simulated data.
    However, large photometric light curve data sets suitable for shape estimation via inversion methods are not typically available for most near-Earth asteroids.
    Therefore for those objects where inversion is not practical, we must understand the additional uncertainty introduced into the interpretation of near-Earth asteroid phase curves that may be impacted by aspect changes throughout or between apparitions.
    
    In this paper, we aim to characterise how aspect changes may affect the interpretation of phase curves for objects where inversion is not possible.
    Section~\ref{sec:hapkeModel} outlines the scattering model and model parameters used throughout this study.
    In Section~\ref{sec:AP10}, we report the detection of aspect change effects on the phase curve of near-Earth asteroid (159402) 1999 AP10, and verify this is an aspect effect through convex inversion to determine the shape and subsequent simulation of the phase curve over the observed apparition.
    In Section~\ref{sec:phaseCurveSimulations}, we discuss simulations of asteroid phase curves over artificial and real asteroid orbits to examine their behaviour and understand the potential magnitude of this effect on other near-Earth asteroids.
    Section~\ref{sec:uncertainty} details software code designed to estimate the aspect-related uncertainty in individual asteroid phase curves based on the geometry over which the asteroids were observed.
    As an example of the significance of this additional uncertainty, we use this software to present updated uncertainties for the phase curve parameters of the near-Earth Asteroids (8014) 1990 MF and (19764) 2000 NF5 originally observed and derived by \citet{2021PASP..133g5003J}.
    In Section~\ref{sec:consequences}, we report on the implications that this effect may have on studies across various wavelength ranges, such as thermophysical modelling.

\section{Hapke Model and Parameters}\label{sec:hapkeModel}
    
    \begin{table}
        \caption{Summary of symbols and parameters used in the Hapke model equations in this paper.}
        \label{tab:params}
        \begin{tabular}{ll}
            \hline
            Parameter       & Explanation \\ \hline\hline
            $i$             & Angle of Incidence  \\
            $e$             & Angle of Reflection \\
            $\psi$          & Azimuth Angle \\
            $\alpha$        & Phase Angle \\
            $\omega$        & Single-Scattering Albedo \\
            $B_0$           & Opposition Effect Amplitude \\
            $h$             & Opposition Effect Width \\
            $g$             & Single-Particle Phase Function Asymmetry \\
            $\bar{\theta}$  &  Global Surface Roughness \\ \hline
        \end{tabular}
    \end{table}

    In this study, we use the Hapke photometric model \citep{hapke2012theory} to simulate the brightness of an asteroid shape model at a given viewing geometry.
    The principle details of the model are reproduced within this section.
    A summary of the symbols and parameters used within the Hapke model equations can be found in Table~\ref{tab:params}.
    The Hapke model parameters assumed in our simulations for the two major taxonomic groups in the near-Earth asteroid regime (S- \& C-type) can be found in Table~\ref{tab:hapkeParams}.
    \begin{table}
        \caption{Hapke photometric model parameters used throughout this study for C-type and S-type asteroids.\newline \textbf{References:} (1) \citet{1989aste.conf..557H}, (2) \citet{2004Icar..172..415L}.}
        \label{tab:hapkeParams}
        \begin{tabular}{lcccccr}
            \hline
            Spec. Type  & $\omega$  & $h$   & $B_0$ & $g$   & $\bar{\theta}$    & Ref.  \\ 
                        &           &       &       &       & (deg)             &       \\ \hline\hline
            C           & 0.037     & 0.025 & 1.03  & -0.47 & 20                & (1)   \\
            S           & 0.330     & 0.010 & 1.40  & -0.25 & 28                & (2)   \\ \hline
        \end{tabular}
    \end{table}

    We calculate the bidirectional reflectance -- the ``ratio of the scattered radiance towards the observer to the collimated incident irradiance'' \citep{2015aste.book..129L} -- of each facet on the surface of the asteroid as
    \begin{eqnarray}\label{eq:rbi}
        r_{bi} = \frac{\omega}{4\pi}\frac{\mu_{0e}}{\mu_e + \mu_{0e}}\left[ (1 + B(\alpha, B_0, h))p(\alpha,g) - \right. \nonumber\\
        \left.1 + H(\mu_{0e}, \omega)H(\mu_e, \omega) \right]S(i,e,\psi,\bar{\theta}).
    \end{eqnarray}
    $\mu_{0e}$ and $\mu_e$ are the effective cosines of the angles of incidence and emission respectively.
    The effective cosines are the mathematical cosines (Eqns. \ref{eq:cosine1} \& \ref{eq:cosine2}) of the angles with a correction applied to account for  global surface roughness \citep{1984Icar...59...41H}.
    \begin{eqnarray}
        &\mu_0& = \cos{i}, \label{eq:cosine1} \\
        &\mu& = \cos{e}. \label{eq:cosine2}
    \end{eqnarray}
    The mathematical formulation of the effective cosines is outlined in Appendix \ref{app:hapke}.
    $B(\alpha,B_0,h)$ is the opposition effect function \citep{1986Icar...67..264H}, which represents the non-linear surge in brightness as an asteroid approaches a phase angle of zero.
    $p(\alpha,g)$ is the single-particle phase function \citep{1941ApJ....93...70H}, which controls the phase dependence of the brightness at higher phase angles away from the opposition effect.
    $H(x, \omega)$ is an approximation that represents multiple scattering \citep{1981JGR....86.3039H}, where incident light is scattered off neighbouring particles at microscopic scales before arriving at the observer location.
    This approximation is found to be valid to within 3 per cent of the numerically evaluated exact form of the \citet{1960ratr.book.....C} H-function over all possible values \citep{1981JGR....86.3039H}, and so is preferred over the numerically evaluated form for simplicity.
    The rough surface shadowing correction $S(i,e,\psi,\bar{\theta})$ describes the modification to the bidirectional reflectance for surface roughness on scales larger than the particle size \citep{hapke2012theory}.
    All of the functions in Eq. \ref{eq:rbi} are defined in full in Appendix \ref{app:hapke}.
    
    The solar flux at the asteroid is calculated by scaling the solar flux at $1\,$au, $S_\lambda = 1.896\,\si{\watt\per\metre\squared\per\nano\metre}$ at $545\,\si{\nano\metre}$ \citep{2004SoEn...76..423G}, by the square of the heliocentric distance of the asteroid, $r$, in au,
    \begin{equation}
        S_{\lambda,r} = \frac{S_{\lambda}}{r^2}.
    \end{equation}
    The facet flux as seen by the observer, $F$, is then calculated as
    \begin{equation}
        F_i = \frac{S_{\lambda,r} \cdot r_{bi} \cdot A \cdot \cos(e)}{\Delta^2},
    \end{equation}
    where $\Delta$ is the geocentric distance of the asteroid in metres, $r_{bi}$ is the bidirectional reflectance as detailed previously ($sr^{-1}$), $A$ is the facet area in square metres, and $e$ is the viewing angle of the facet (equivalent to the reflection angle of the observed ray from the facet).
    Illumination by the Sun and visibility to the observer (accounting for shadowing by other facets) of each facet of the shape model is checked using geometry calculations originally implemented as part of the ATPM \citep{2011MNRAS.415.2042R}.
    The flux from each facet is calculated for and summed over all facets that are both illuminated and visible, then converted into a reduced magnitude in the Johnson V waveband as
    \begin{equation}
        V(1,1,\alpha) = -2.5\log_{10}{\frac{\Sigma_{i}F_i}{V_0}} - 5\log_{10}{r\Delta},
    \end{equation}
    where $V_0 = 3.631 \times 10^{-11}\,\si{\watt\per\metre\squared\per\nano\metre}$ is the zero-point flux in the Johnson V-band \citep{1998A&A...333..231B}, and $r$ and $\Delta$ are in au.
    We can therefore simulate the reduced magnitude of any defined shape at a given geometry over each step in its light curve to obtain the expected average reduced magnitude at each point on its phase curve.
    
\begin{figure}
        \includegraphics[width=\linewidth]{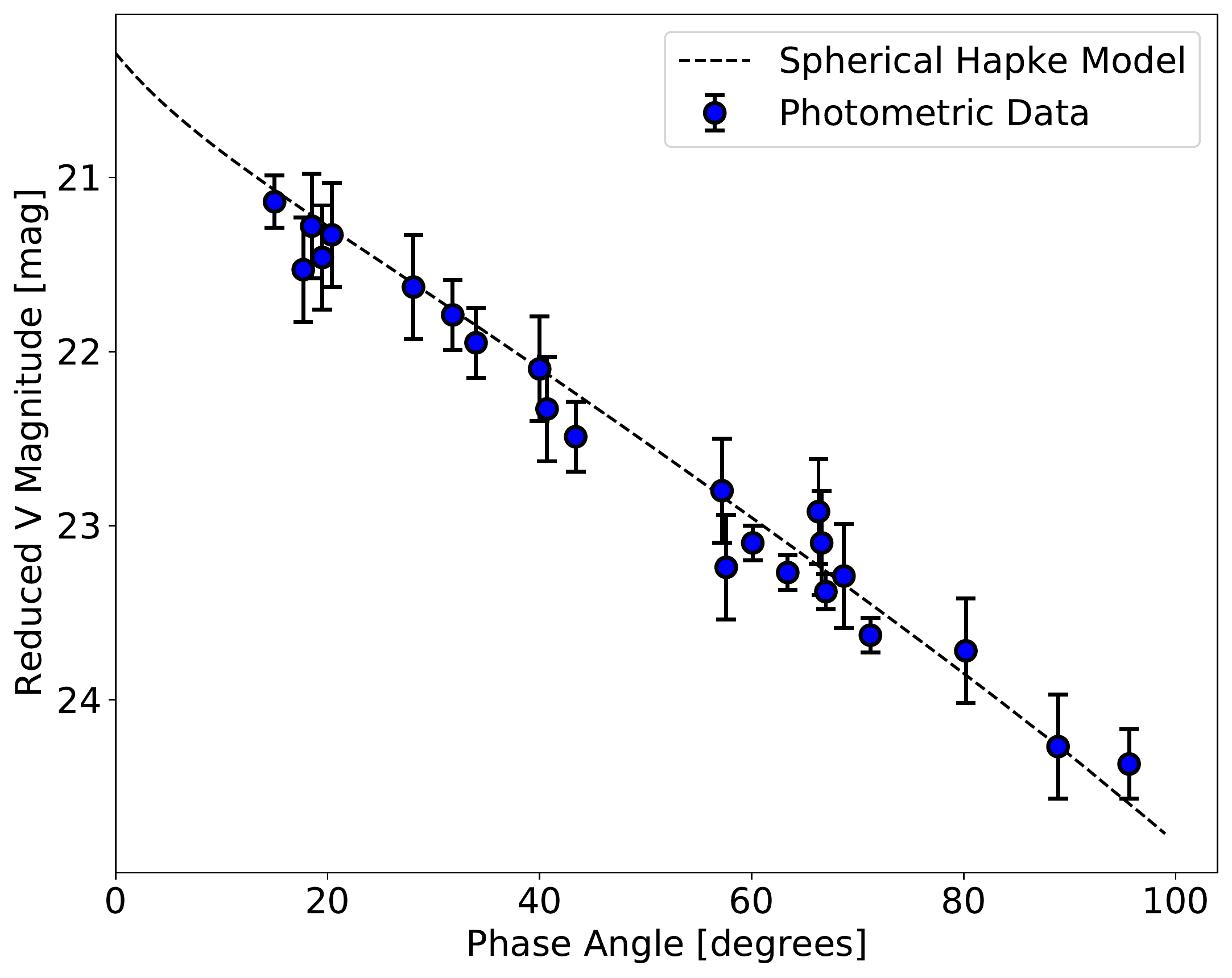}
        \caption{Simulated phase curve of OSIRIS-REx target (101955) Bennu using Hapke parameters derived by \citet{2015Icar..252..393T}, verifying our implementation of the Hapke model through comparison with the photometric data from \citet{2013Icar..226..663H}. Error bars on data points correspond to one-sigma uncertainties.}
        \label{fig:bennuVerification}
    \end{figure}

    We verify the performance and accuracy of our implementation of the Hapke photometric model through simulation of the phase curve of the OSIRIS-REx target (101955) Bennu.
    This is achieved through comparison of the photometric data from \citet{2013Icar..226..663H} to a spherical model simulated using the Hapke scattering parameters derived by \citet{2015Icar..252..393T}.
    This model phase curve, over-plotted with the photometric data, is shown in Figure~\ref{fig:bennuVerification}.
    The simulated phase curve well matches the photometric data and the Hapke model presented in Figure 4 of \citet{2015Icar..252..393T}, verifying the implementation of the model.
    
\section{Convex Shape Modelling and Phase Curve Variability of (159402) 1999 AP10}\label{sec:AP10}
    
\begin{figure}
        \centering
        \includegraphics[width=\linewidth]{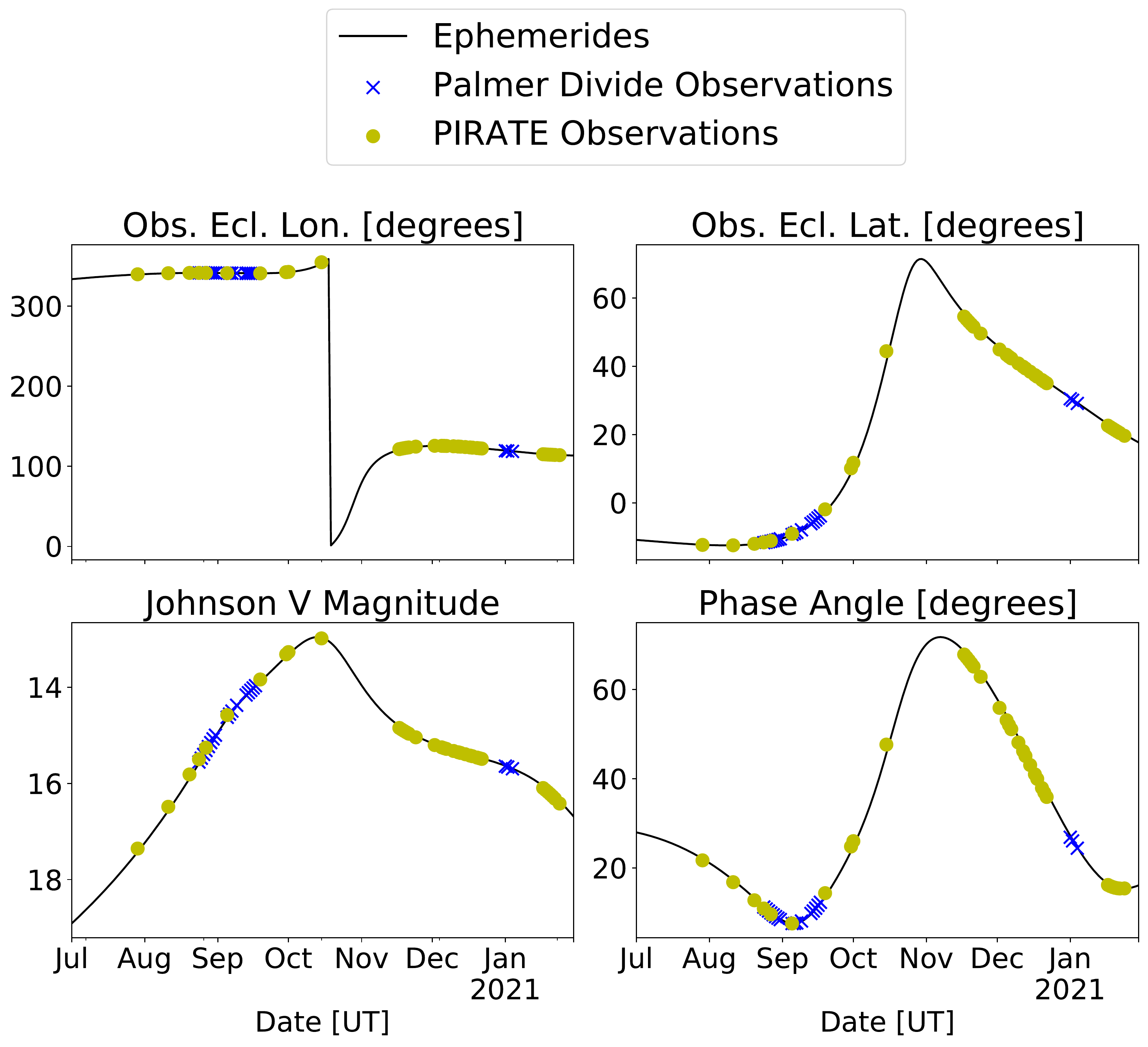}
        \caption{Observing geometry of (159402) 1999 AP10 during observations in 2020 and 2021. The top panels show the position of the asteroid in the ecliptic coordinate system as viewed from Earth. The bottom two panels show the change in brightness and phase angle of the asteroid throughout the apparition. The black line represents the ephemerides, the blue crosses represent the geometry at which the Palmer Divide observations were taken, and the yellow dots represent the geometry at which the PIRATE observations were taken.}
        \label{fig:AP10obsGeom}
    \end{figure}

    As part of an ongoing near-Earth asteroid observation program with the Physics Innovations Robotic Telescope Explorer \textbf{\citep[PIRATE; described in][]{2018RTSRE...1..127K,2021PASP..133g5003J}}, the NEA (159402) 1999 AP10 was observed over 38 nights from July 2020 - January 2021 with \textasciitilde 212 hours of photometric data.
    The data were reduced and calibrated according to the processes set out by \citet{2021PASP..133g5003J}.
    The observational coverage with PIRATE is marked by the yellow points over-plotted on top of the ephemerides (black line) in Figure \ref{fig:AP10obsGeom}.
    The asteroid was also observed during this apparition at Palmer Divide Observatory (PDO) by \citet{2021MPBu...48...40W}. The light curve data were obtained from the Asteroid Lightcurve Data Exchange Format (ALCDEF) database\footnote{\url{https://minplanobs.org/alcdef/}} \citep{2011MPBu...38..172W}.
    The observational coverage of the PDO data are marked by the blue crosses in Figure \ref{fig:AP10obsGeom}.
    
    (159402) 1999 AP10 is a near-Earth asteroid with an Amor orbit classification.
    Radar, thermal-IR, and many light curve studies have been conducted on this asteroid to date.
    \citet{2018PASJ...70..114H} derive a synodic rotation period of $7.911 \pm 0.0001$ hours, in agreement with the synodic period of $7.908 \pm 0.001$ hours from \citet{2010MPBu...37...83F}.
    This is also consistent with more recent observations by \citet{2021MPBu...48...40W}, who report synodic rotation periods of $7.9219 \pm 0.0003$ hours and $7.9144 \pm 0.0004$ hours.
    The small differences in derived periods across these studies is largely due to these being synodic rather than sidereal rotation periods, i.e. the intrinsic rotation period is modified by the changing viewing geometry over the observations.
    \citet{2021MPBu...48...40W} also report the potential detection of a secondary rotation period from their light curve observations, although the secondary period is not well constrained and is significantly different when using different data sets.
    To investigate whether a secondary body is present, we viewed four hours of delay-Doppler radar data\footnote{Available at: \url{https://www.lpi.usra.edu/resources/asteroids/asteroid/?asteroid_id=1999AP10}.} \citep{LPIradarAP10} and find no evidence of a secondary.
    This means that either the secondary was behind the main body, is too small to be detected at this radar resolution ($30\,\si\metre\,$px$^{-1}$), is out of frame, or is not present.
    For the purposes of this study, we do not consider the presence of a binary, and any such binary would be too small to influence the conclusions presented in this work (contributing $\leq 0.004\,$ mag to the observed magnitude).
    We also find no evidence of a secondary in the $75\,\si\metre$-resolution Goldstone delay-Doppler data\footnote{Available at: \url{https://echo.jpl.nasa.gov/asteroids/1999AP10/1999ap10.2020.planning.html}.}, which provides another geometry to help constrain the presence or absence of a secondary.
    
    Warm Spitzer observations give a diameter estimate of $1.20 \pm 0.29\, \si{\kilo\metre}$ \citep{2010AJ....140..770T,2011AJ....141..109M}, inconsistent with a radar-derived diameter of $2.068\,\si{\kilo\metre}$ \citep{LPIradarAP10}.
    Since radar data provide a more direct measurement of the sizes of solar system bodies, we prefer this diameter in our analysis.
    From the warm Spitzer data the following additional parameters are derived: a geometric albedo of $0.35 \pm 0.24$ \citep{2011AJ....142...85T}, and an Sq classification \citep{2014Icar..228..217T} in the Bus-DeMeo taxonomic system.
    Spectroscopic data from the MIT-Hawaii Near-Earth Object Spectroscopic Survey suggests an Sw classification \citep{2019Icar..324...41B}, whereas other spectroscopic results suggest either an L- or S-type classification \citep{2009ATel.2323....1H,2009ATel.2283....1H}.
    The circular polarisation ratio from the radar continuous wave spectrum \citep[SC/OC = $0.26$;][]{LPIradarAP10} is consistent with an S-type classification \citep{2008Icar..198..294B}.
    
    \begin{figure}
        \centering
        \includegraphics[width=\linewidth]{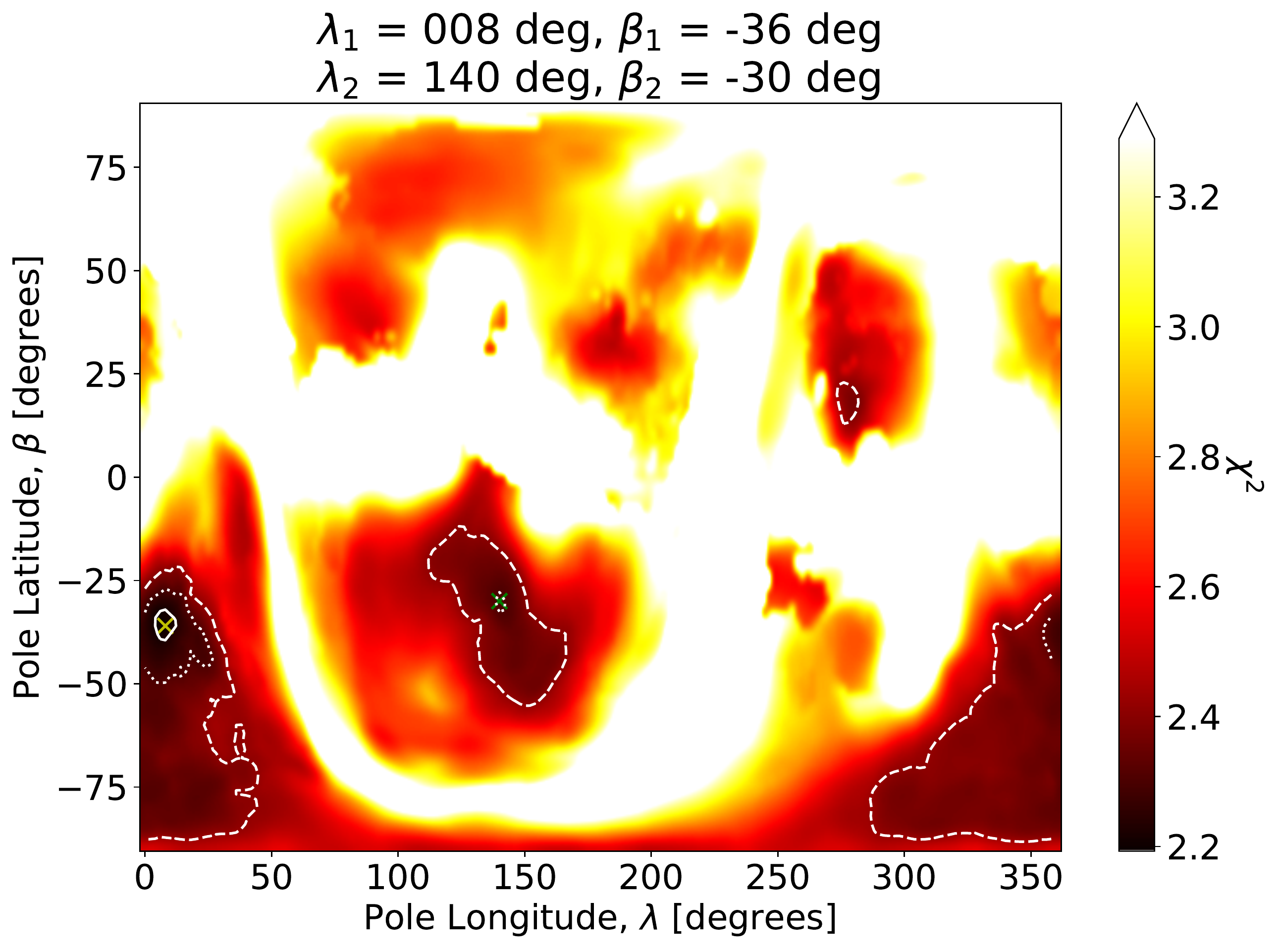}
        \caption{Pole scan for (159402) 1999 AP10, represented as a $\chi^2$ plane indicating goodness of fit for each possible pole solution in a $2\si\degree \times 2\si\degree$ grid. The best-fit pole solution is indicated by the yellow cross at $\lambda = 8\pm4\si\degree$, $\beta=-36\pm2\si\degree$, with a $\chi^2$ value of $2.193$. The $1$ per cent, $5$ per cent, and $10$ per cent uncertainty contours are represented by the solid, dotted, and dashed lines respectively. Pole solutions $50$ per cent above the minimum $\chi^2$ are marked by the white area of the plot. A second potential pole solution within the $5$ per cent contour is indicated by the green cross at $\lambda = 140\pm4\si\degree$, $\beta=-30\pm2\si\degree$, with a $\chi^2$ value of $2.298$.}
        \label{fig:AP10pole}
    \end{figure}
    \begin{figure*}
        \centering
        \includegraphics[width=\linewidth]{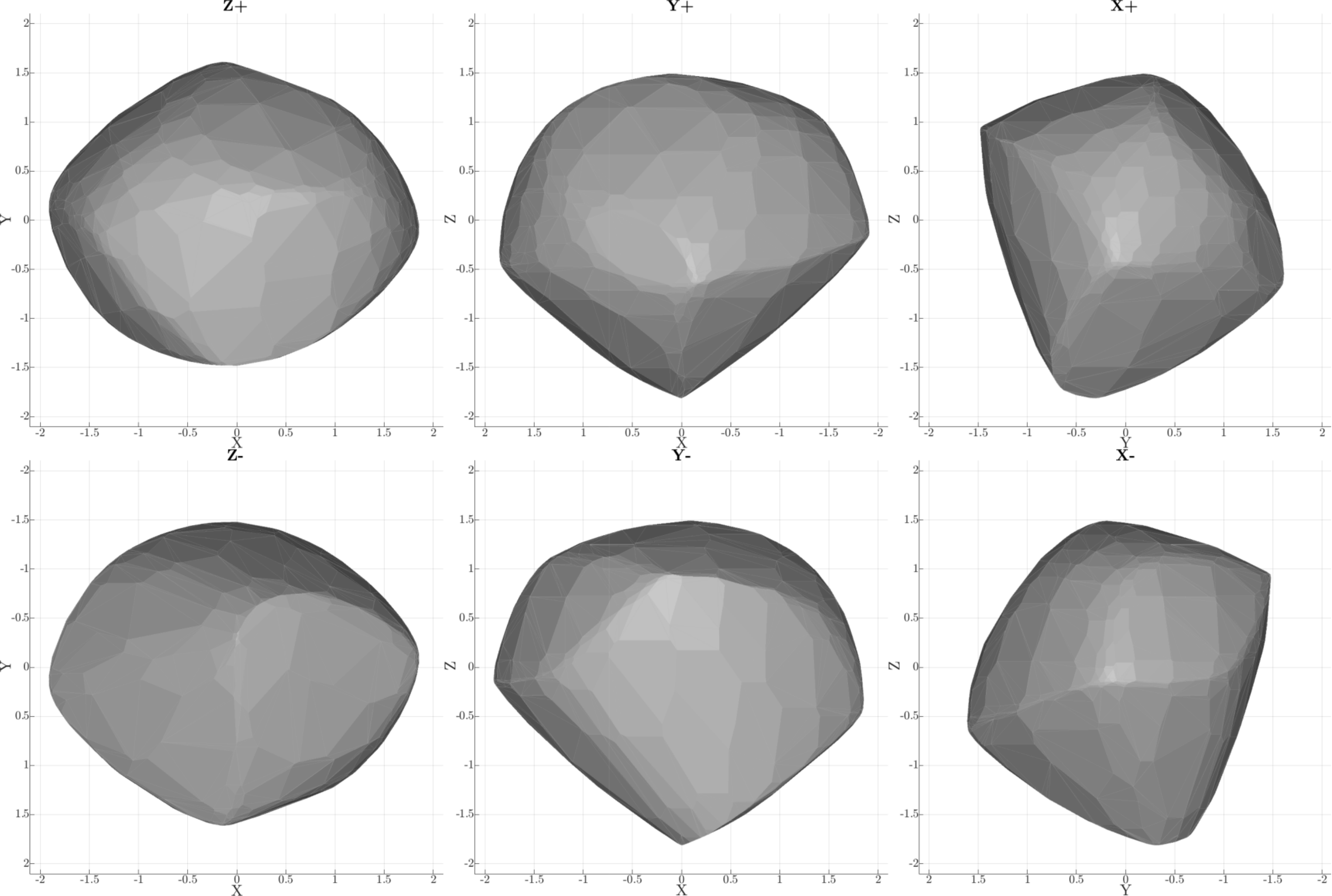}
        \caption{The best-fit convex shape model of (159402) 1999 AP10, with pole coordinates $\lambda = 8\si\degree$, $\beta = -36\si\degree$, and sidereal rotation period $P = 7.9214 \pm 0.0008$ hours. The rotation pole is aligned with the Z-axis, with the body-centric north in the positive Z direction. \textbf{Top row:} views along the $Z$, $Y$, and $X$ axes of the shape-centric coordinate system. \textbf{Bottom row:} views along the Z, Y, and X axes from the other side compared to the top row. Axis scales do not correspond to any physical quantities.}
        \label{fig:AP10model}
    \end{figure*}
    \begin{figure*}
        \centering
        \includegraphics[width=.83\linewidth]{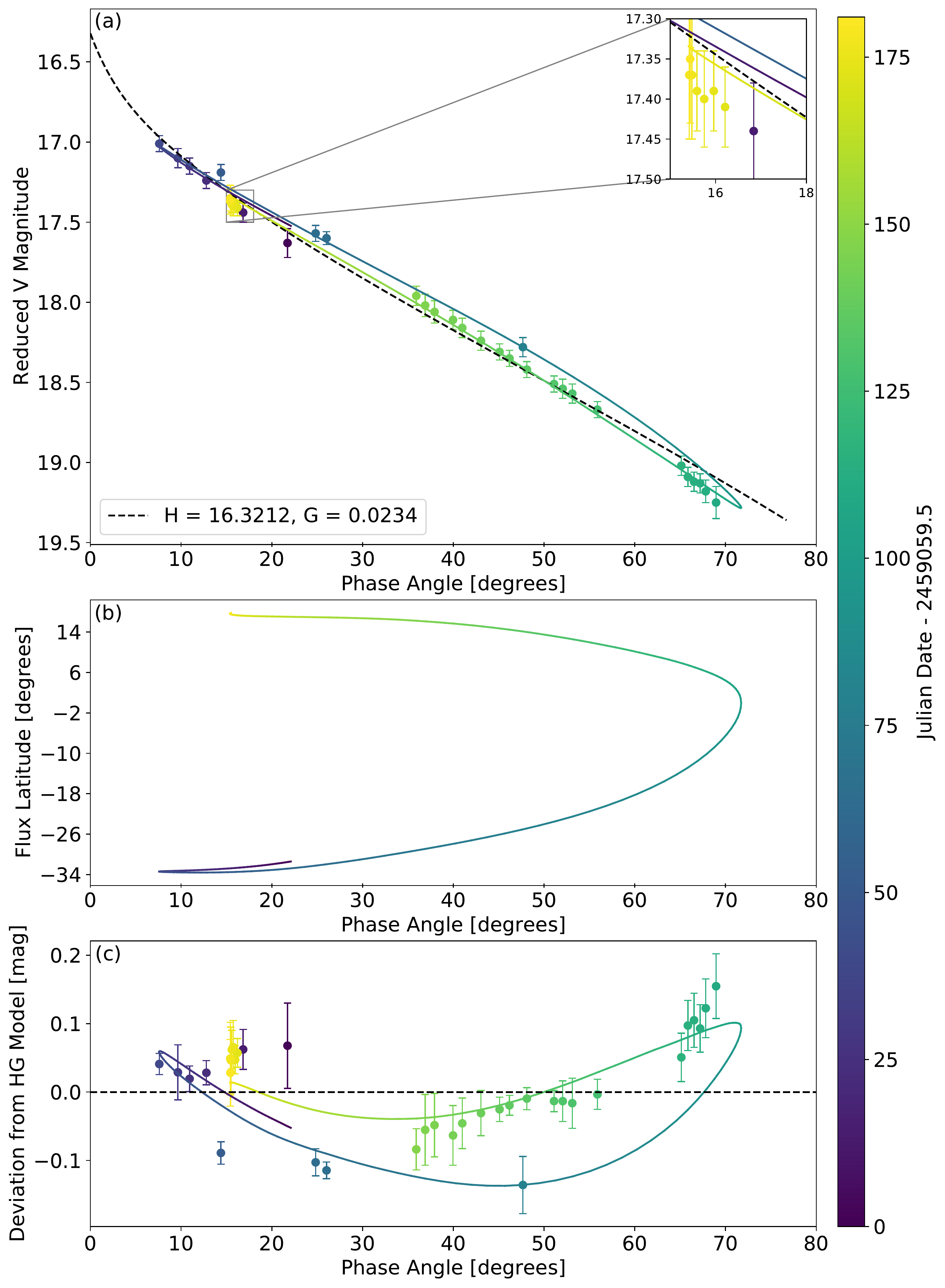}
        \caption{\textbf{(a)} Phase curve data for (159402) 1999 AP10 (individual points). The solid line (which bends twice) is the simulated phase curve over the apparition using the derived convex shape model and a Hapke photometric model. S-type Hapke parameters are assumed for this object. Colour corresponds to the Julian Date. Error bars on each data point include a nightly systematic calibration uncertainty, and are therefore larger than typical one-sigma uncertainties. Dashed black line corresponds to the best-fit $HG$ model to the phase curve data ignoring aspect effects. \textbf{(b)} The `flux latitude', defined as the flux-weighted mean latitude of illuminated facets visible by the observer, plotted against phase angle. Colour corresponds to the Julian Date as in the upper panel. Over the apparition, the predominant source of the observed flux changes from the southern hemisphere of the asteroid to the northern hemisphere. This hemispherical switch triggers the aspect-related deviations of the phase curve in the upper panel due to the shape differential between the hemispheres of this asteroid. \textbf{(c)} Deviations from the best-fit $HG$ model of the observed phase curve data (circles) and the simulated phase curve using the shape model from the best-fit $HG$ model (solid line). In this panel the systematic calibration uncertainties have been removed to leave the error bars as typical one-sigma uncertainties.}
        \label{fig:AP10phaseCurve}
    \end{figure*}
    
    After the observing run with PIRATE, a phase curve was constructed from the data.
    It was observed that different parts of the apparition produced different slopes on the phase curve.
    Due to the rapidly changing viewing geometry during the apparition it was proposed that this variation in the phase curve could be due to shape effects at different viewing aspects.
    To investigate this possibility, relative light curves from PIRATE were combined with the PDO light curves to produce a convex shape model.
    To achieve this, the \textsc{convexinv} software package was used \citep{2010A&A...513A..46D}, which aims to approximate the shape and pole orientation of an asteroid through convex inversion techniques \citep{2001Icar..153...24K,2001Icar..153...37K}.
    
    To model the shape, a pole scan is performed following procedures established in \citet{2019A&A...627A.172R,2019A&A...631A.149R} and \citet{2021MNRAS.507.4914Z}.
    This involves setting up a grid of possible pole positions covering the entire celestial sphere in a $2\si\degree \times 2\si\degree$ grid.
    An initial estimate of the sidereal rotation period came from a \textsc{convexinv} period scan of the photometric data, searching for solutions consistent with the previously reported synodic rotation periods.
    For each pole in the grid, the shape and period is optimised using \textsc{convexinv} to minimise the deviation between the model and observed light curves.
    The $\chi^2$ statistic between the observed light curves and the model light curves from the optimised shape is recorded for each pole, to create a 2-D $\chi^2$ plane as shown in Figure \ref{fig:AP10pole}.
    The pole (and corresponding optimised convex shape and sidereal rotation period) with the minimum $\chi^2$ value is taken as the best solution.
    The yellow cross in the figure represents the minimum $\chi^2$ solution.
    The solid contour encompasses solutions within $1$ per cent of the minimum $\chi^2$, the dotted contour encompasses those within $5$ per cent, and the dashed encompasses those within $10$ per cent.
    The uncertainty on the pole solution beyond the $1$ per cent contour is quite large due to the limited observational coverage over one apparition.
    
    This gives a pole solution for this object at $\lambda_1 = 8 \pm 4 \si\degree$, $\beta_1 = -36 \pm 2 \si\degree$, with a sidereal rotation period of $7.9214 \pm 0.0008 \si\hour$ (uncertainties correspond to $1$ per cent contours in the $\chi^2$ plane), and with $\chi^2 = 2.193$.
    A second potential pole solution within the $5$ per cent contour is found at $\lambda_2 = 140\pm4\si\degree$, $\beta_2 = -30\pm2\si\degree$, with a $\chi^2$ value of $2.298$.
    For the purposes of this study we only consider the best pole solution, and state the second for reference.
    There exist additional light curve data from the last bright apparition in 2009, although these data cover an almost identical range of viewing geometries and are therefore of limited use in further constraining the pole.
    The period is also already constrained well enough to justify not including the 2009 data.
    The optimised convex shape is shown in Figure \ref{fig:AP10model}, and the model light curves are shown plotted with the real data in Figure S1 in the supplementary online material.
    The sidereal rotation period derived in this work is consistent with the synodic periods reported previously, and the retrograde pole solution is consistent with predictions of retrograde rotation by \citet{2021MPBu...48...40W} due to the change in synodic rotation period over the apparition.
    
    Once the shape model for this asteroid was obtained, we simulated the reduced magnitude of the asteroid over its apparition using the shape model and S-type parameters in the Hapke photometric model described in Section~\ref{sec:hapkeModel}.
    As the phase curve uses rotationally averaged data, a convex model is sufficient to reconstruct the data and therefore a radar shape model is not required for this level of analysis.
    The reduced magnitudes simulated using the model are plotted on the phase curve alongside the calibrated photometric data from PIRATE in Figure \ref{fig:AP10phaseCurve}(a).
    The colour of the points and the line corresponds to the date of observation, with the colour changing from purple to yellow as the apparition progressed from 2020-07-29 to 2021-01-24.
    The simulated phase curve accurately reconstructs the observed data and associated deviations over the apparition.
    The deviations of both the data and the simulated phase curve from the best-fit $HG$ model are plotted in Figure \ref{fig:AP10phaseCurve}(c), helping to demonstrate the divergence from a typical phase curve.
    In these figures, the simulated data is shifted vertically by a small amount to compensate for differences between the assumed effective albedo in the model (from the assumed S-type parameters) and the actual albedo of this asteroid.
    This shift also compensates for uncertainty in the radar-derived diameter.
    
    As an explanation of the source of the aspect-related deviations over the phase curve, we plot in Figure~\ref{fig:AP10phaseCurve}(b) a quantity we define as the `flux latitude' ($\psi_\text{flux}$)  against phase angle, with colour corresponding to Julian Date as before.
    The flux latitude is defined as the flux-weighted mean latitude of illuminated shape model facets visible by the observer.
    This tells us from which latitude the flux is predominantly being observed, i.e. it can tell us whether we are mainly `seeing' the northern or southern hemisphere of the shape.
    We see that over the apparition, the flux latitude changes by approximately 50 degrees.
    At the beginning of the apparition the observer is mainly seeing flux from the southern hemisphere of this asteroid, but as the apparition progresses this shifts towards the northern hemisphere.
    The hemispherical asymmetry in the shape of this object therefore triggers aspect-related deviations in the phase curve.

\section{Aspect Change Simulations}\label{sec:phaseCurveSimulations}

    \subsection{Artificial Viewing Geometries}\label{subsec:unrealisticGeoms}
        
\begin{figure}
            \centering
            \begin{subfigure}{\linewidth}
                \includegraphics[width=\linewidth]{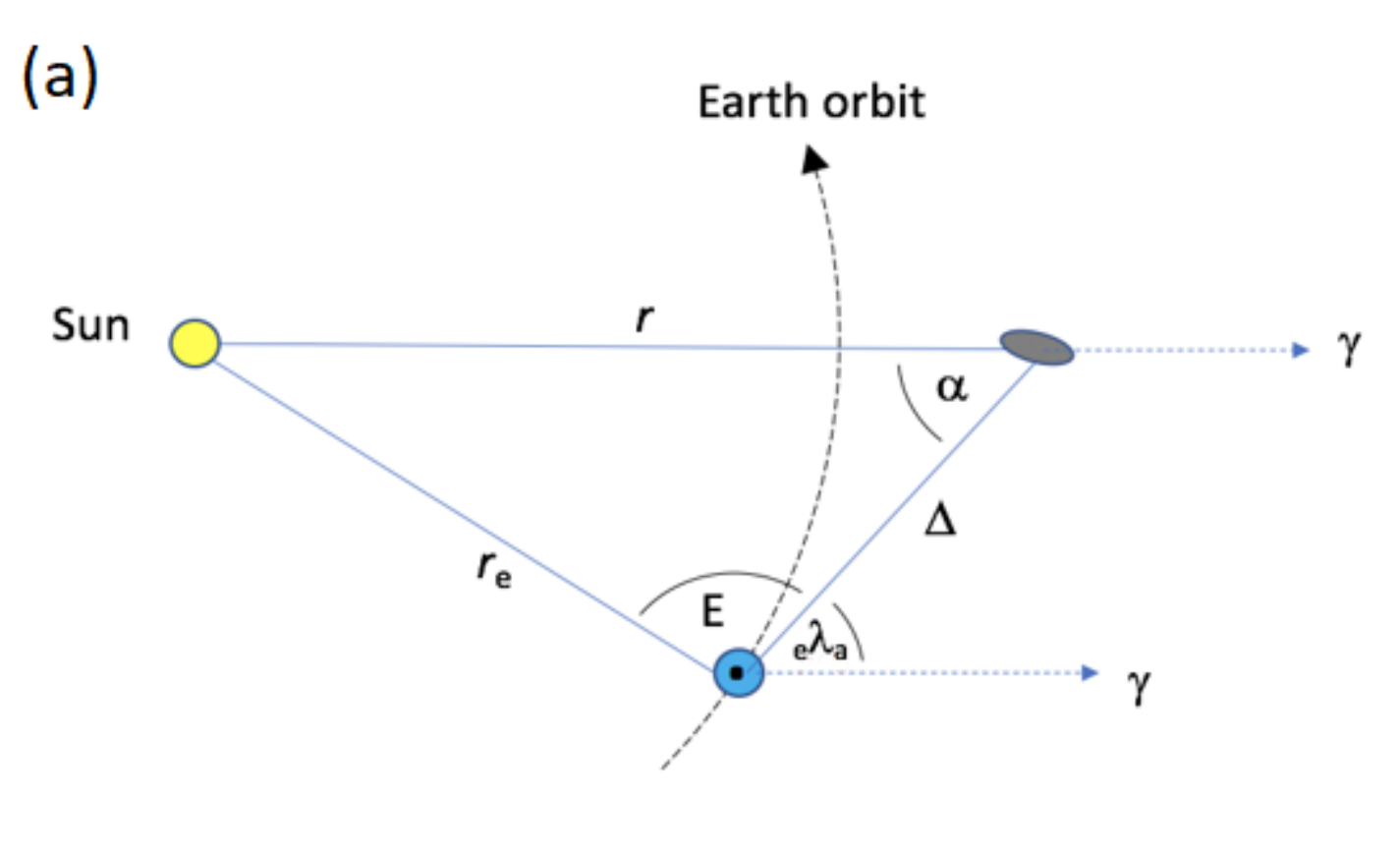}
            \end{subfigure}
            \begin{subfigure}{\linewidth}
                \includegraphics[width=\linewidth]{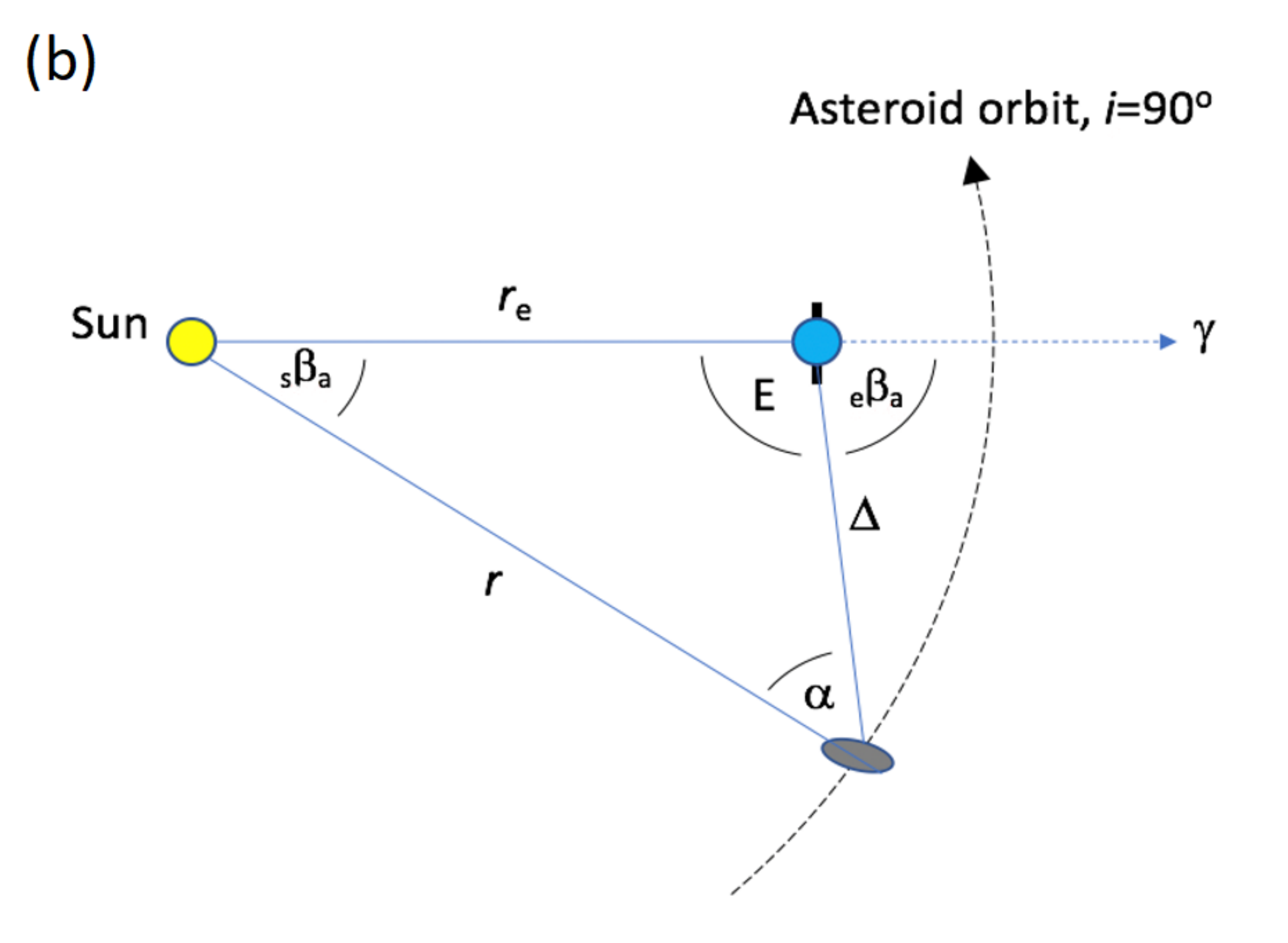}
            \end{subfigure}
            \caption{\textbf{(a)} Orientation 1 - asteroid position is held fixed with $r = 1.0001$\,au, while Earth sweeps inside. \textbf{(b)} Orientation 2 - Earth position is held fixed, while asteroid sweeps from south to north with $r = 1.0001$\,au. Panel (b) partially reproduced from Figure 1 in \citet{2021EPSC...15..255J} under a Creative Commons Attribution 4.0 License.}
            \label{fig:simulationGeometries}
        \end{figure}

        Following our detection of variability in the phase curve of (159402) 1999 AP10, an investigation was conducted into the potential effects this may have on the interpretation of the large collection of photometric data available for near-Earth asteroids.
        The scale at which this effect may occur and how it may manifest itself under various circumstances must be characterised through simulation in order to understand the potential effect in photometric observations.
        To achieve this, 21 ellipsoidal shape models were created from 2048 facets with: a 1 km volumetric diameter, various ellipsoid parameters and pole orientations.
        A single spherical model is also created as a control (aspect changes will have no effect on the phase curve of a spherical object).
        The parameters and ecliptic pole coordinates of these ellipsoids are detailed in Table~\ref{tab:ellipsoids}.
        These ellipsoid models were simulated over two artificial, but useful, viewing geometries.
        The first involved keeping the asteroid stationary at a heliocentric distance $r = 1.0001\,$au while the Earth sweeps by interior to the orbital distance of the asteroid (Figure~\ref{fig:simulationGeometries}a, referred to herein as orientation 1).
        The second geometry assumes the Earth is fixed and the asteroid, with an orbital inclination $i = 90^\circ$ and $r = 1.0001\,$au, sweeps from South to North  (Figure~\ref{fig:simulationGeometries}b, referred to herein as orientation 2).
        Both of these geometries allow for a pre- and post-opposition phase curve to be simulated with a phase angle range covering $90\si\degree$ -- $0\si\degree$ -- $90\si\degree$.
        As before with (159402) 1999 AP10, all reduced magnitudes were simulated using the Hapke photometric model with S-type parameters and rotationally averaged.
        
        \begin{table}
            \caption{22 ellipsoidal models (including spherical control) used to simulated phase curves over the geometries in Figure \ref{fig:simulationGeometries}. $\lambda$ and $\beta$ are rotation pole coordinates in the ecliptic coordinate system. $a$, $b$, $c$ are the principle semi-axes of the ellipsoid. $c_1$ and $c_2$ are used to introduce hemispherical asymmetry into the ellipsoid model \citep{1989Icar...78..298C}.}
            \label{tab:ellipsoids}
            \begin{tabular}{lccrrrr}
                \hline
                Model No.   & $\lambda$ & $\beta$   & $a$   & $b$   & $c_1$ & $c_2$ \\ 
                            & (deg)     & (deg)     &       &       &       &       \\ \hline\hline
                0           & 0         & 90        & 1     & 1     & 1     & 1     \\
                1           & 0         & 90        & 2     & 1     & 1     & 1     \\
                2           & 0         & 90        & 2     & 1.5   & 1     & 1     \\
                3           & 0         & 90        & 2     & 1.5   & 1.25  & 0.75  \\
                4           & 0         & 45        & 2     & 1     & 1     & 1     \\
                5           & 0         & 45        & 2     & 1.5   & 1     & 1     \\
                6           & 0         & 45        & 2     & 1.5   & 1.25  & 0.75  \\
                7           & 0         & 0         & 2     & 1     & 1     & 1     \\
                8           & 0         & 0         & 2     & 1.5   & 1     & 1     \\
                9           & 0         & 0         & 2     & 1.5   & 1.25  & 0.75  \\
                10          & 45        & 45        & 2     & 1     & 1     & 1     \\
                11          & 45        & 45        & 2     & 1.5   & 1     & 1     \\
                12          & 45        & 45        & 2     & 1.5   & 1.25  & 0.75  \\
                13          & 45        & 0         & 2     & 1     & 1     & 1     \\
                14          & 45        & 0         & 2     & 1.5   & 1     & 1     \\
                15          & 45        & 0         & 2     & 1.5   & 1.25  & 0.75  \\
                16          & 90        & 45        & 2     & 1     & 1     & 1     \\
                17          & 90        & 45        & 2     & 1.5   & 1     & 1     \\
                18          & 90        & 45        & 2     & 1.5   & 1.25  & 0.75  \\
                19          & 90        & 0         & 2     & 1     & 1     & 1     \\
                20          & 90        & 0         & 2     & 1.5   & 1     & 1     \\
                21          & 90        & 0         & 2     & 1.5   & 1.25  & 0.75  \\ \hline
            \end{tabular}
        \end{table}
        
\begin{figure*}
            \centering
            \begin{subfigure}{.33\textwidth}
                \includegraphics[width=\textwidth]{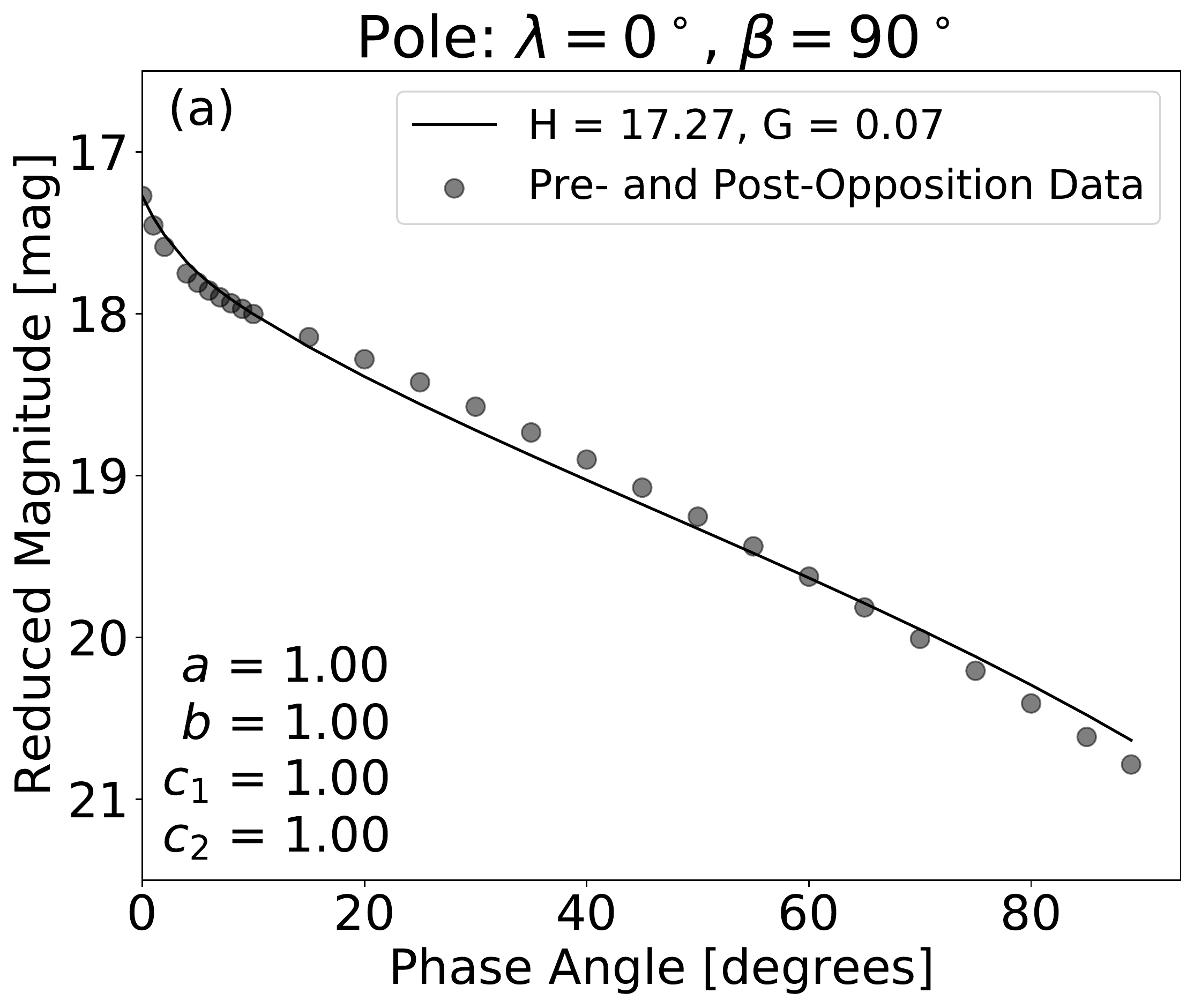}
            \end{subfigure}
            \begin{subfigure}{.33\textwidth}
                \includegraphics[width=\textwidth]{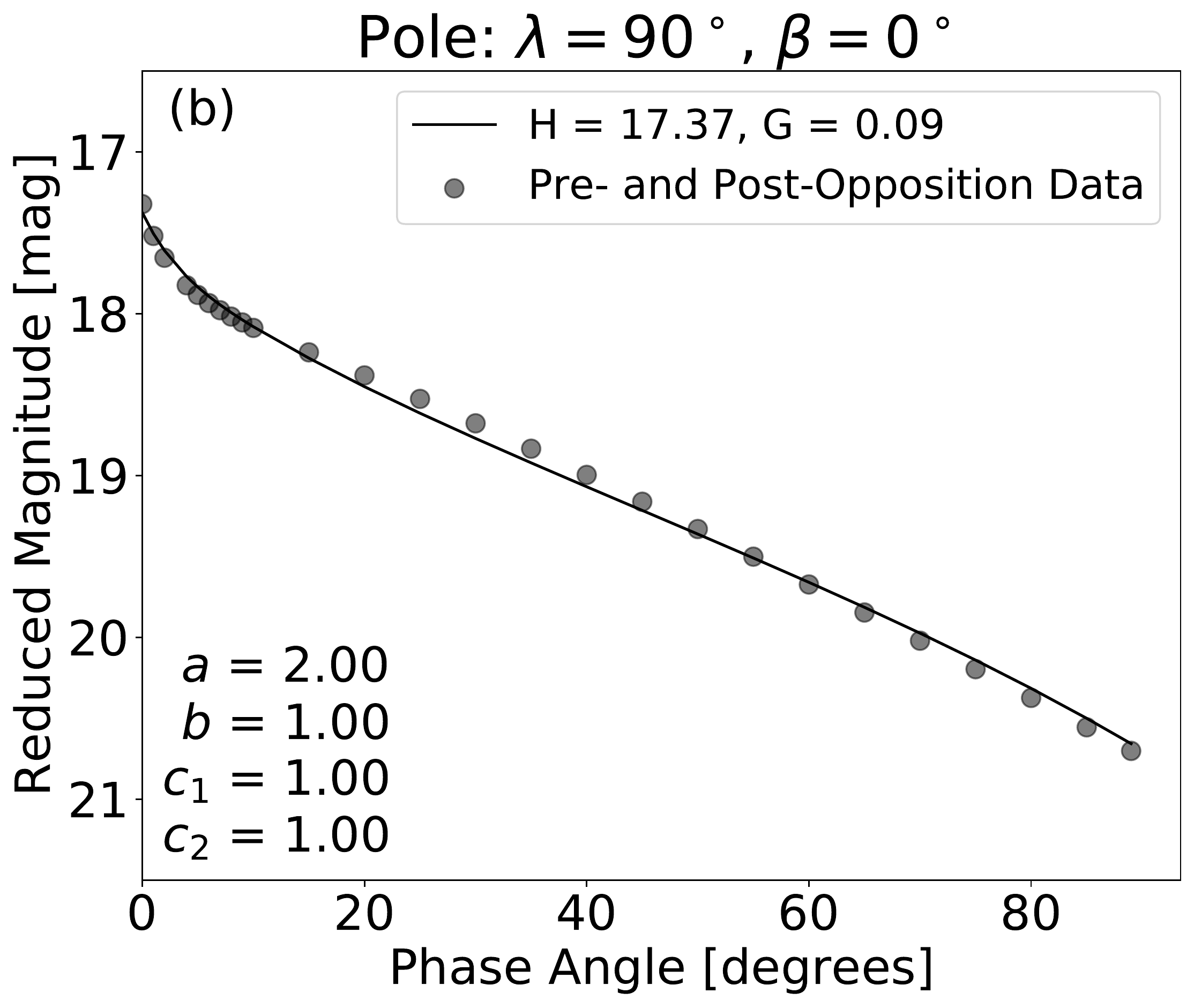}
            \end{subfigure}
            \begin{subfigure}{.33\textwidth}
                \includegraphics[width=\textwidth]{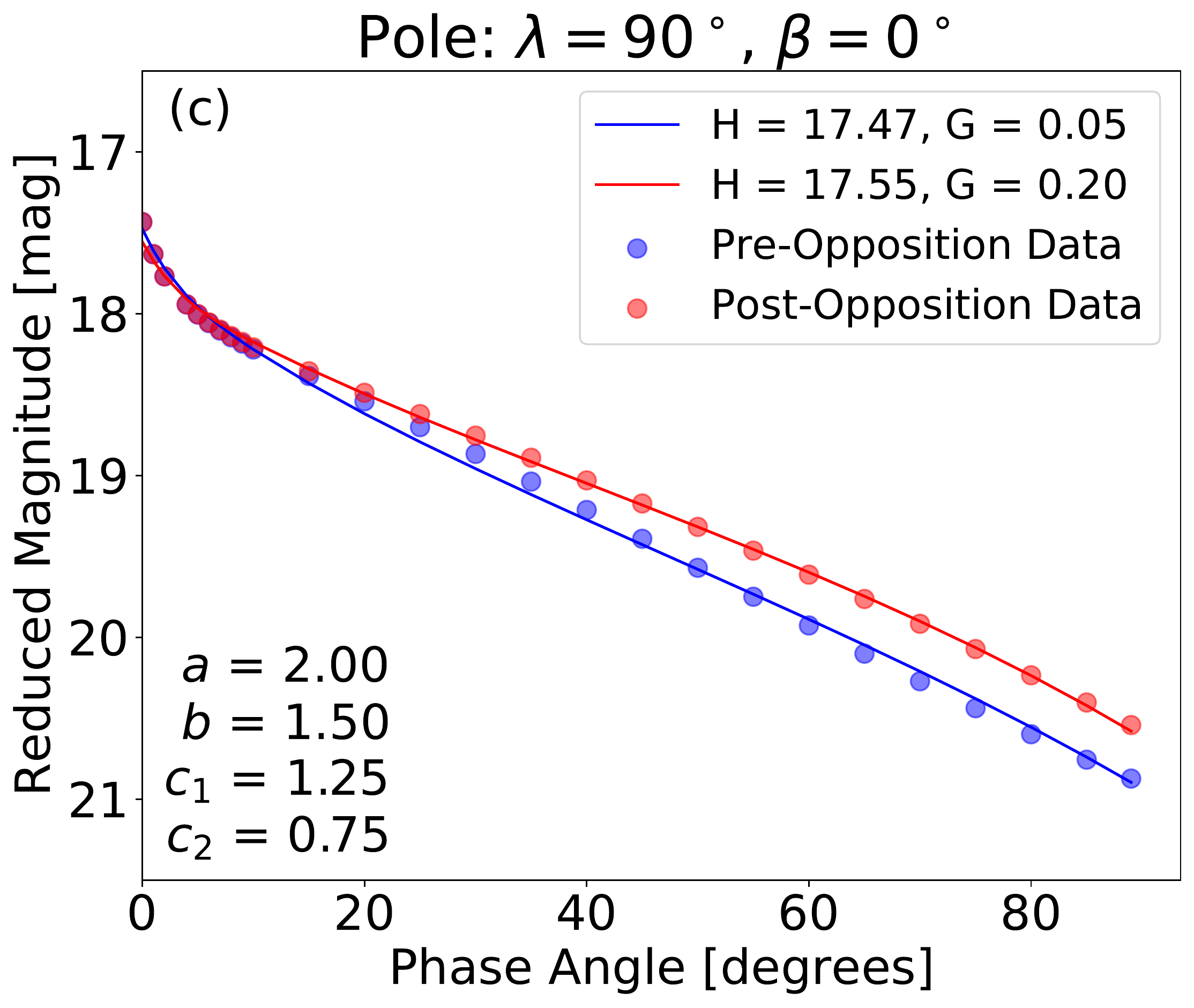}
            \end{subfigure}
            \caption{Simulations for orientation 1. \textbf{(a)} Spherical control model. \textbf{(b)} Ellipsoidal model with elongated `a' axis, with a pole orientation in the ecliptic. The phase curve deviates from the spherical control. \textbf{(c)} Ellipsoidal model with hemispherical asymmetry, with a pole orientation in the ecliptic. The phase curve not only deviates from the spherical control, but deviates between pre- and post-opposition simulated data.}
            \label{fig:orientation1_sample}
        \end{figure*}

        In the case of orientation 1, a selection of three models are shown in Figure \ref{fig:orientation1_sample}.
        All of the simulations over this geometry can be found in Figure S2 in the supplementary online material.
        Each sub-figure contains the simulated phase curve data and a fit to the data using the $H$, $G$ photometric system.
        The fitted $H$ and $G$ parameters are displayed in the legend of each figure.
        The first model is the spherical control, which shows the expected $G$ parameter for the S-type Hapke parameters simulated.
        The second model is an elongated object with its rotation pole in the ecliptic, resulting in a phase curve which deviates from that of the spherical control model.
        The third model is elongated, and introduces hemispherical asymmetry in the same way as \citet{1989Icar...78..298C}, with a pole orientation in the ecliptic once again.
        This results not only in further deviations from the spherical control, but deviations between the pre- and post-opposition phase curves for the same object.
        This indicates that not only can phase curves deviate for different shapes despite having the same surface scattering properties, but the same object can show different phase curves within a single apparition.
        These results are consistent with similar simulations of asteroid phase curves by \citet{1992LIACo..30..353K}.

\begin{figure*}
            \centering
            \begin{subfigure}{.33\textwidth}
                \includegraphics[width=\textwidth]{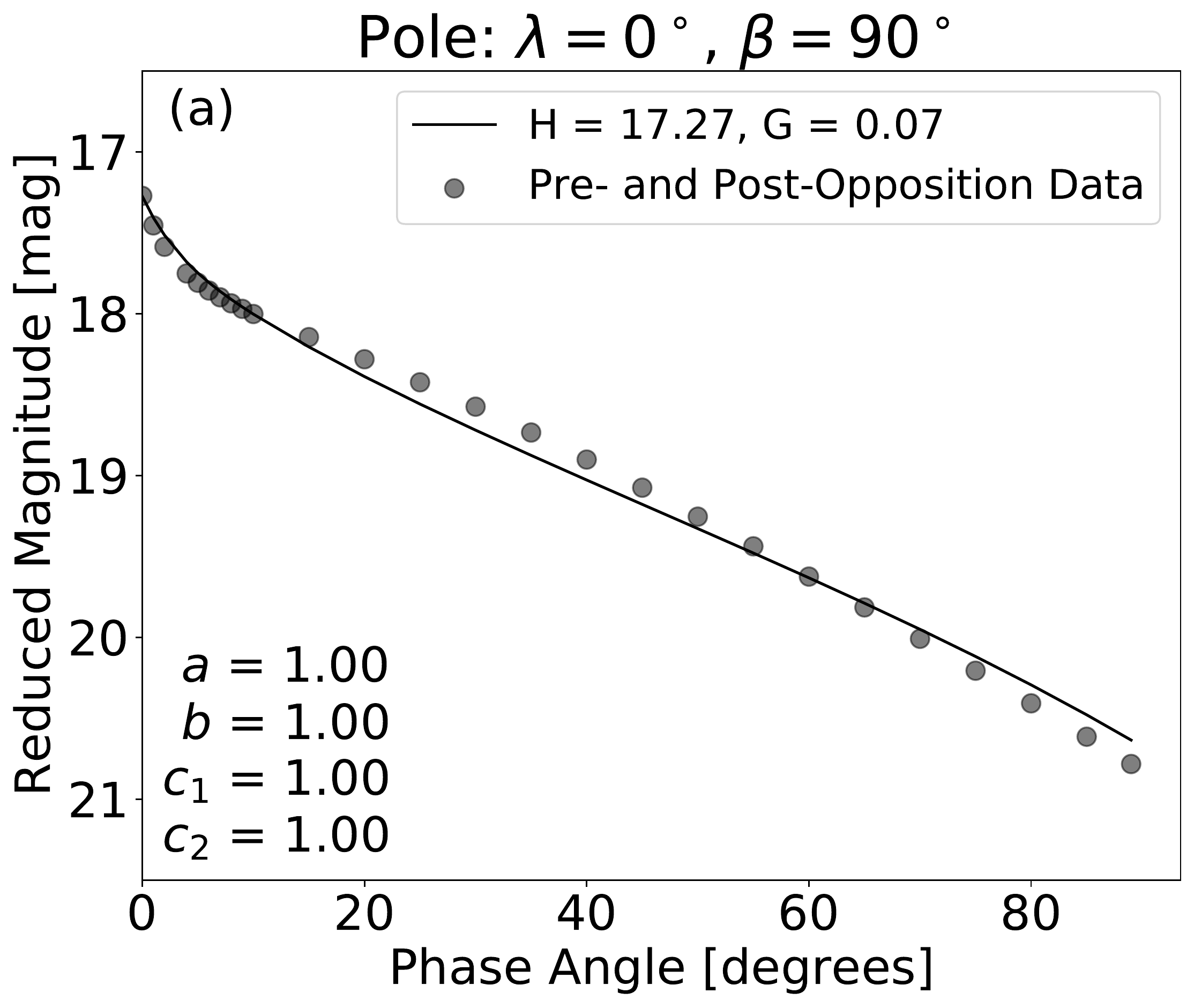}
            \end{subfigure}
            \begin{subfigure}{.33\textwidth}
                \includegraphics[width=\textwidth]{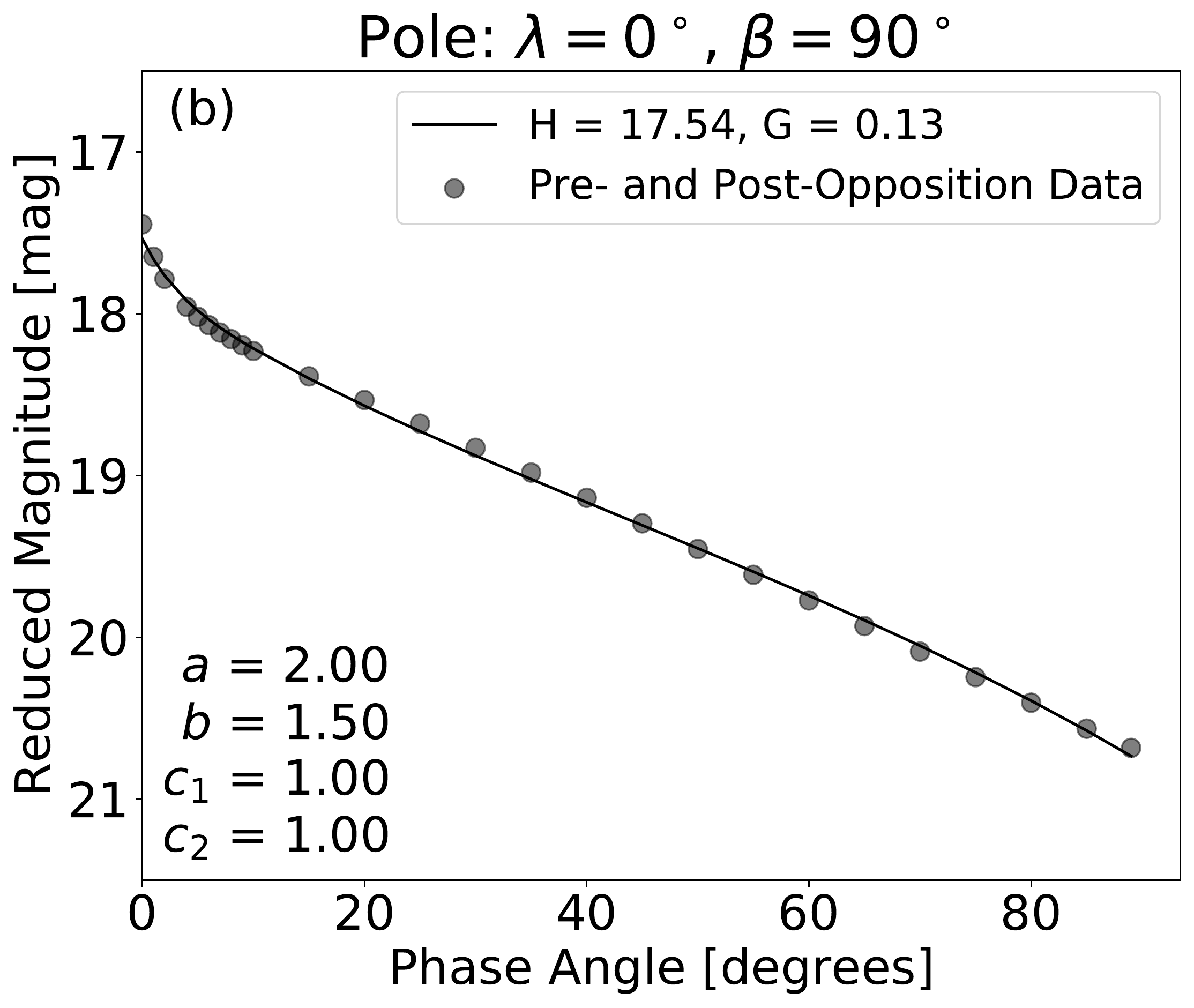}
            \end{subfigure}
            \begin{subfigure}{.33\textwidth}
                \includegraphics[width=\textwidth]{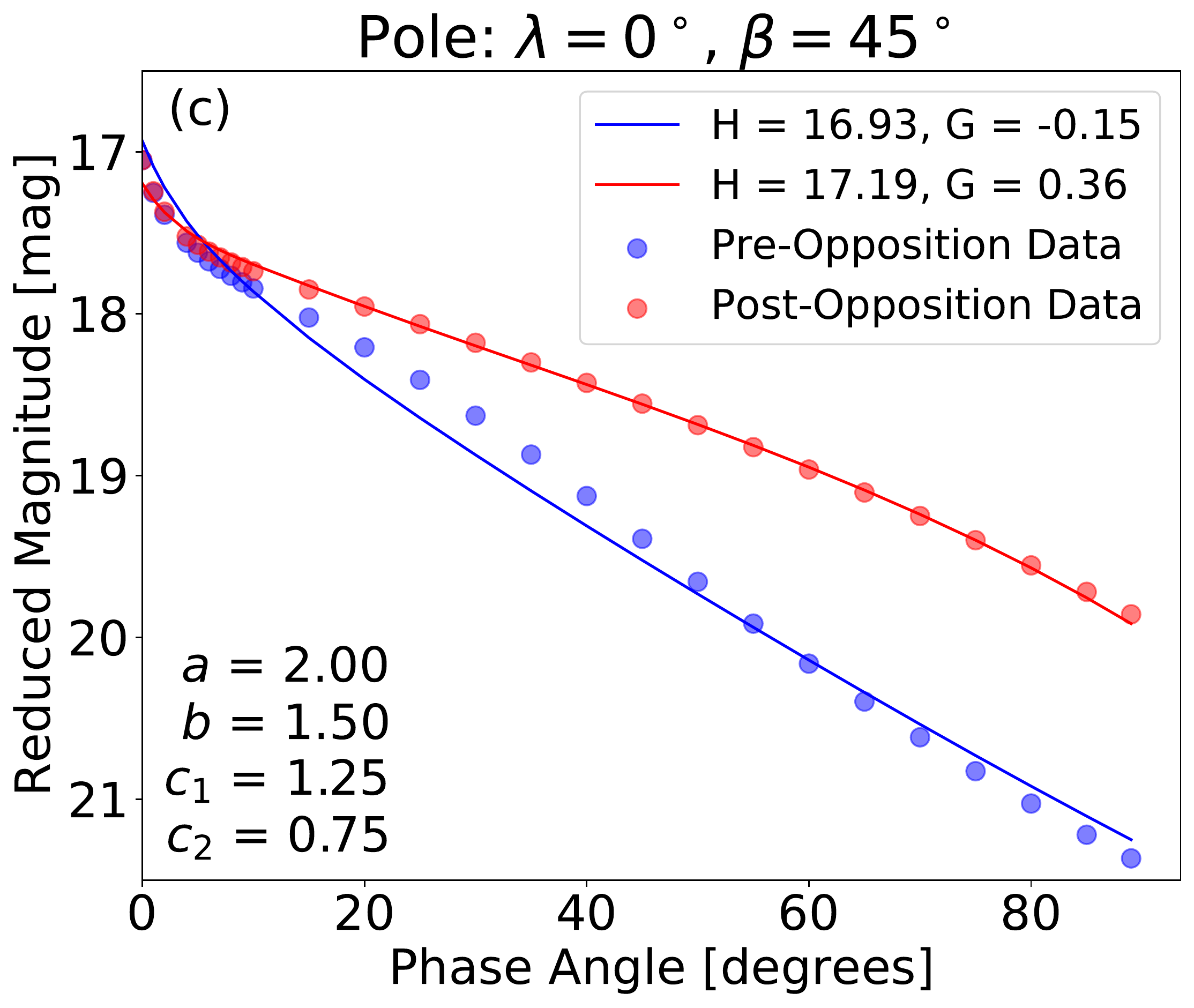}
            \end{subfigure}
            \caption{Simulations for orientation 2. \textbf{(a)} Spherical control model. \textbf{(b)} Ellipsoidal model with elongated `a' and `b' axes, with a pole orientation perpendicular to the ecliptic. The phase curve deviates from the spherical control. \textbf{(c)} Ellipsoidal model with hemispherical asymmetry, with a pole orientation angled $45\si\degree$ to the ecliptic. The phase curve not only deviates from the spherical control, but deviates between pre- and post-opposition simulated data. Figure partially reproduced from Figure 2 in \citet{2021EPSC...15..255J} under a Creative Commons Attribution 4.0 License.}
            \label{fig:orientation2_sample}
        \end{figure*}

        A sample of three models from orientation 2 is shown in Figure \ref{fig:orientation2_sample}.
        All of the simulations over this geometry can be found in Figure S3 in the supplementary online material.
        Once again the spherical control is the first model, and is consistent with that from orientation 1 as expected.
        The second model again introduces elongation and puts the pole orientation perpendicular to the ecliptic, resulting in a deviation from the spherical control.
        The third model (elongated, hemispherical asymmetry, and pole 45 degrees to the ecliptic) shows a drastic difference between pre- and post-opposition phase curves.
        As an example of how much this may affect existing methodology for interpreting individual asteroid phase curves, we attempt taxonomic classification of the simulated object based on the two phase curves for pre- and post-opposition in Figure \ref{fig:orientation2_sample}c using the single parameter taxonomic fitting method from \citet{2016P&SS..123..117P}.
        Using this method we obtain: a C-type classification for the pre-opposition phase curve, and an E-type classification for the post-opposition phase curve.
        The timing of when the asteroid is observed during the apparition drives the taxonomic classification using this method in this example.
        This effect therefore severely limits the taxonomic information present in near-Earth asteroid phase curves.
        Care must be taken when trying to extract this information from phase curves, and should typically be supported by spectroscopic or spectro-photometric observations.
        The large-scale separation of the phase curve at higher phase angles in Figure \ref{fig:orientation2_sample}c highlights the importance of this effect for near-Earth asteroid phase curves, where higher phase angle data is more common than for main belt asteroids.
        
\begin{figure*}
            \centering
            \includegraphics[width=\textwidth]{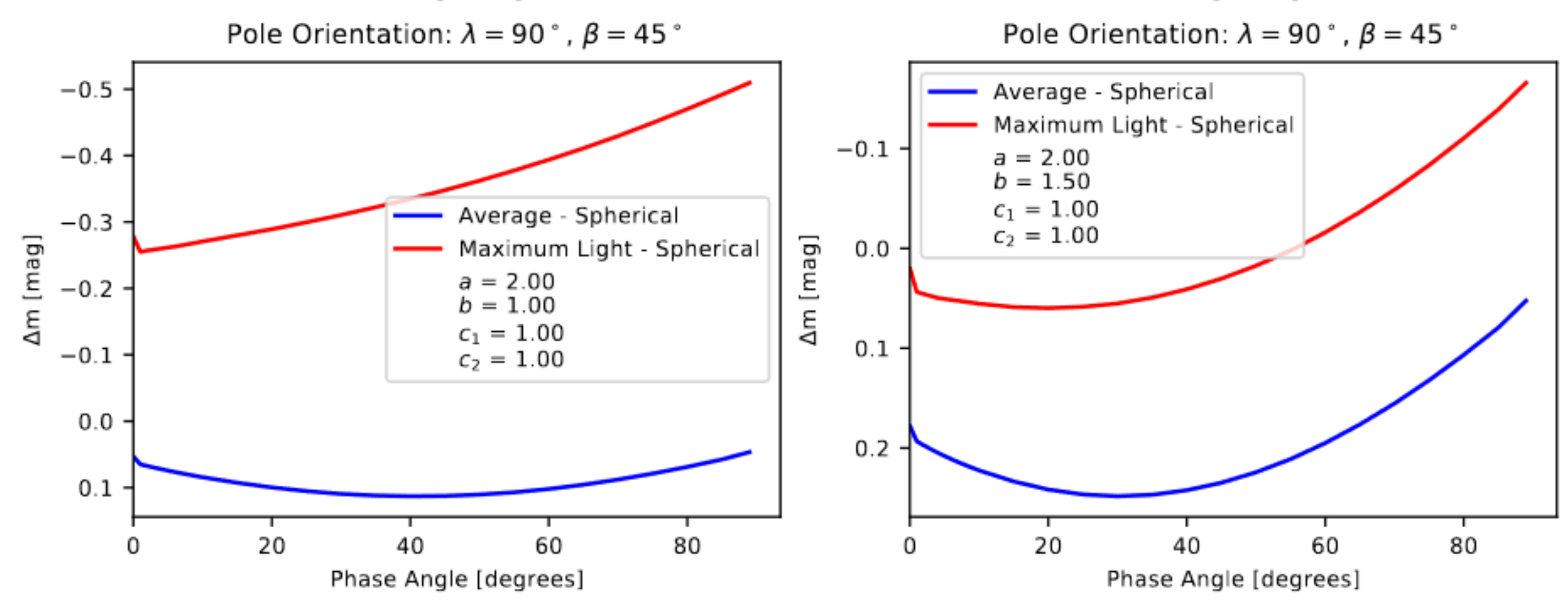}
            \caption{\textbf{Left Panel:} Difference between the averaged phase curve and the spherical control (blue), and difference between the maximum brightness phase curve and the spherical control (red). Simulated asteroid is elongated by the ellipsoid $a$ parameter. In this case the average light phase curve provides the best approximation to the spherical control. \textbf{Right Panel:} Same as the left panel, however the shape of the simulated asteroid is altered slightly in the $b$ parameter. In this case the maximum brightness phase curve provides the best approximation to the spherical control. Whether or not the maximum brightness phase curve improves the approximation to the `real' phase curve is highly variable and can change with the slightest variation in shape of the asteroid, and is therefore not a viable solution to the problems posed by aspect changes.}
            \label{fig:sphericalComps}
        \end{figure*}

        A potential solution to minimise the effect of aspect changes on phase curves was thought to be to define the phase curves not as the \textit{average} brightness variation with phase angle, but instead to define them as the \textit{maximum} brightness variation with phase angle.
        For each of the simulations and geometries described previously, we compared the average brightness phase curve to the spherical control (the `true' phase curve for the selected Hapke parameters).
        We also compared the maximum brightness phase curves to the spherical control.
        In these simulations we see no evidence that the maximum brightness phase curve is better in general at reproducing the spherical control.
        Simulations of maximum brightness asteroid phase curves by \citet{1992LIACo..30..353K} result in similar effects to those shown in Fig.~\ref{fig:orientation1_sample}c, and hence these effects are not removed through the use of maximum brightness phase curves.
        In Figure \ref{fig:sphericalComps}, we present two example scenarios where we have compared the average brightness phase curve to the spherical control, and similarly compared the maximum brightness phase curve to the spherical control.
        These two simulations provide a stark example of how changing one parameter in the simulation (here it is the $b$ ellipsoid parameter) can change whether the average brightness or maximum brightness phase curves provide a better approximation to the spherical control.
        Using maximum brightness phase curves therefore does not provide an adequate solution to the problem, and phase curves should continue to be constructed using rotationally averaged data as per standard practice \citep{1989aste.conf..557H}.
        
    \subsection{Simulating Over Real Asteroid Geometries}\label{subsec:realGeoms}
        
        \begin{figure}
            \centering
            \includegraphics[width=\linewidth]{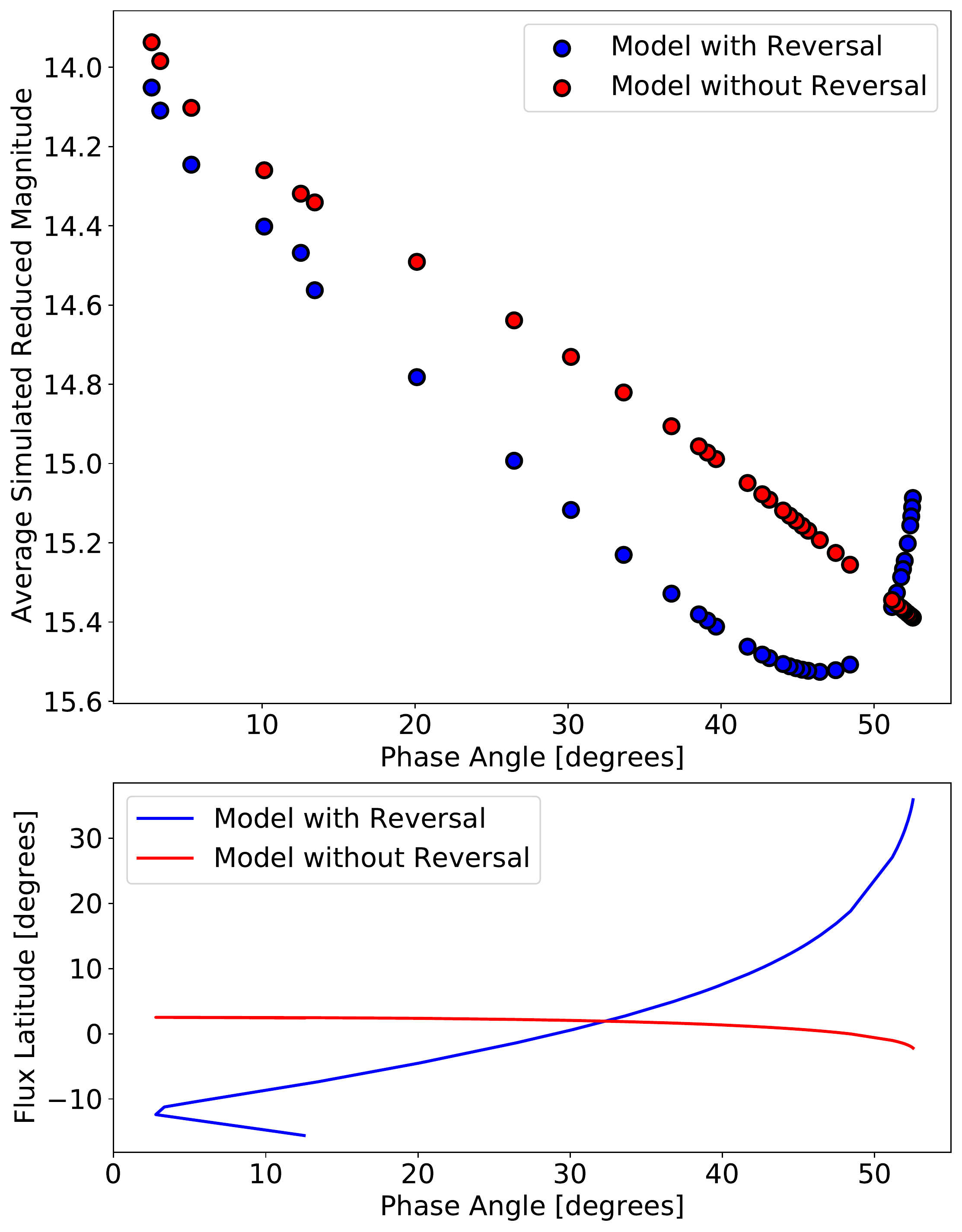}
            \caption{\textbf{Upper Panel:} The red points are an example of a simulation of the phase curve of an ellipsoidal model over the orbit of (19764) 2000 NF5 that does not undergo a reversal in the phase curve. The blue points are an example of a simulation of the phase curve of an ellipsoidal model over the orbit of (19764) 2000 NF5 that undergoes a reversal against the typical phase curve direction due to aspect effects. Data collected at these points in the phase curve can be significantly misleading in the analysis of the phase curve parameters derived from these data. \textbf{Lower Panel:} Change in flux latitude over the simulated apparition for both models. For the model with phase curve reversal, the change in the flux latitude is significant over the apparition, which leads to the aspect effects we see in the phase curve. For the case with no observed reversal in the phase curve, the flux latitude is not observed to change rapidly, and hence aspect effects are not as clearly visible in the simulated phase curve.}
            \label{fig:reversalVnoReversal}
        \end{figure}
        
        \begin{figure*}
            \centering
            \includegraphics[width=\textwidth]{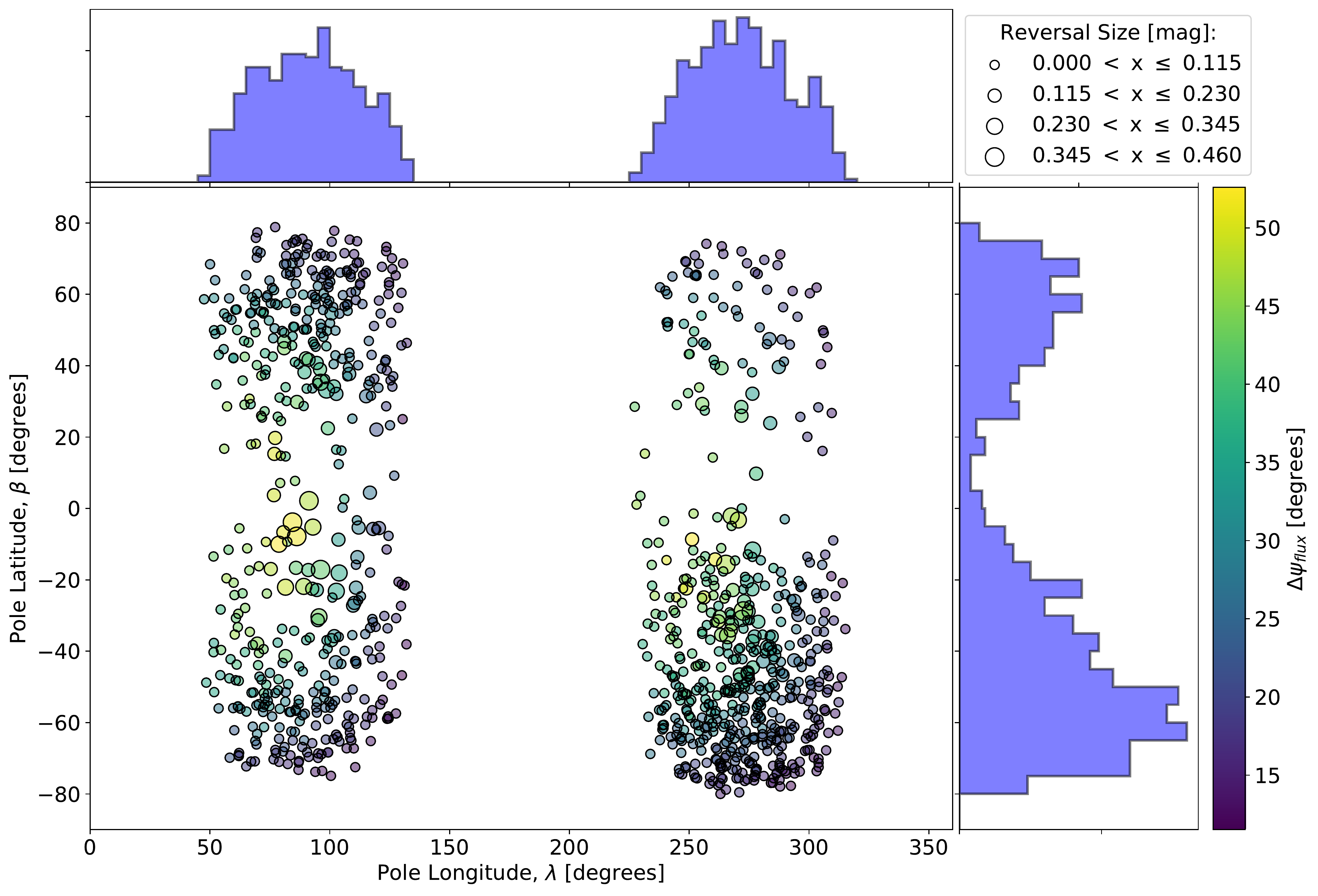}
            \caption{Pole orientations of models which showed a reversal in the phase curve when simulated over the 2020 apparition of (19764) 2000 NF5. Sizes of the individual points correspond to the magnitude of the reversal in the phase curve, with the largest reversals occurring with pole orientations close to the ecliptic. Distribution of the pole latitudes largely follows that of the input distribution, except for an absence of reversals with poles perpendicular to the ecliptic. Distribution of pole longitudes deviates strongly from the uniform input distribution, with only two groups of pole longitudes able to create reversals in the phase curves.}
            \label{fig:19764reversals}
        \end{figure*}
        
        While it is useful to visualise the worst-case scenarios that may be observed in near-Earth asteroid phase curves, the geometries described previously are not representative of typical orbits.
        As an example of how this effect may present itself during an apparition of a real NEA, we investigate the effect over the apparition of (19764) 2000 NF5 from June to October 2020.
        To do this, four thousand random ellipsoidal models were generated with random pole orientations.
        In the absence of an unbiased NEA shape distribution in the literature, the ellipsoidal models were randomly chosen from uniform distributions on the ellipsoid parameters ($a$, $b$, $c_1$, and $c_2$), while maintaining $a > b$ and $a,\,b > c_1,\,c_2$.
        The pole orientations were sampled using a uniform distribution in longitude, and an obliquity distribution from \citet{2017A&A...608A..61T}.
        We then simulated the phase curve of each test object using the Hapke photometric model and S-type parameters described previously.
        It was noted that in \textasciitilde $26.6$ per cent of simulations, the phase curve underwent a reversal in direction at or around a phase angle of $50\si\degree$, i.e. the reduced magnitude began to decrease with larger phase angle.
        These reversals arise solely from aspect changes over the apparition.
        Due to the changes in aspect, the cross-sectional area illuminated and visible by the observer rapidly changed.
        This imparted brightness variations on the simulated phase curve that dominate over the expected reduction in brightness due to typical phase curve behaviour of a single consolidated body, hence we see an increase in brightness on the simulated phase curve with increasing phase angle.
        
        An example of one of these simulations where the phase curve reversed in direction can be found in Figure \ref{fig:reversalVnoReversal} (blue points, upper panel).
        An example of one of the simulations where there was no reversal, i.e. the phase curve showed `normal' behaviour, is illustrated by the red points in the same figure.
        In the lower panel of Figure \ref{fig:reversalVnoReversal}, we plot the flux latitude against phase angle to confirm the cause of the reversal in the phase curve.
        For the reversal case, the flux latitude changes rapidly at the same time as the reversal in the phase curve, indicating we are rapidly sampling different parts of the shape.
        There is minimal change in the flux latitude for the model that did not undergo any reversal in the phase curve.
        This shows that it is the aspect-driven brightness variations that cause these reversals in the phase curve as described previously.
        This demonstrates the large effect that changing aspect can have on phase curves.
        
        To analyse under what scenarios the reversals in the phase curves occurred across all models, we plotted the pole coordinates of all of the test objects that provided a reversal (Figure \ref{fig:19764reversals}).
        The relative lack of reversals with the pole in the ecliptic is a result of the input distribution instead of any geometric effects over the simulated apparition.
        The pole longitudes are focused in two regions, between $50\si\degree$ - $120\si\degree$ and between $235\si\degree$ - $310\si\degree$.
        This is significantly different from the uniform input distribution, and indicates that under this geometry the phase curves are heavily dependent on the pole coordinates.
        The largest reversals in the phase curve occurred with the pole latitude near to the ecliptic, which over this geometry results in large scale changes to the flux latitude and hence large aspect-related effects in the phase curves.
        The overall distribution of pole latitudes does not deviate significantly from the input distribution, except for the lack of poles perpendicular to the ecliptic where hemispherical asymmetry has no effect under this geometry.

\section{Estimating Aspect Related Uncertainty in Phase Curve Parameters}\label{sec:uncertainty}

    We note in the previous sections that the magnitude of aspect effects is highly dependent on viewing geometry and pole orientation.
    This means that a generalised approach to estimate a single aspect-related uncertainty in all asteroid phase curves is not possible.
    We have developed software written in \textsc{c++} and \textsc{Python} that aims to estimate this uncertainty for asteroid phase curves on an individual basis for asteroids for which no information on shape, pole orientation or spectral type is available.
    Using the same process outlined in Section~\ref{subsec:realGeoms}, one thousand ellipsoidal models are generated with pole orientations according to the outlined input distributions.
    The models are all scaled such that the fluxes are equivalent to an object with a $1\,\si{\kilo\meter}$ volumetric diameter.
    
    The user provides an input geometry file containing: the heliocentric range, heliocentric longitude, heliocentric latitude, geocentric range, geocentric longitude, geocentric latitude, and photometric uncertainty of the asteroid on each date on which it was observed in the real phase curve.
    The uncertainties are used to maintain consistency between constraining the best fit parameters for each simulation and for the real photometric data.
    The phase curve of the object is then simulated over this geometry using both S- and C-type Hapke scattering parameters, using the model and parameters outlined in Section~\ref{sec:hapkeModel}.
    We fit the $H$, $G$ parameters to each simulated phase curve using non-linear least squares fitting implemented in the \textsc{SciPy} python package \citep{2020NatMe..17..261V}.
    The standard deviation in each of the fitted phase curve parameters across all 1000 models is recorded as the aspect-related uncertainty for that parameter, $\sigma_{H\text{, aspect}}$ \& $\sigma_{G\text{, aspect}}$.
    This process is done for models with both S- and C-type Hapke scattering parameters.
    S-type simulations show greater aspect effects due to the increased multiple scattering present on these surfaces.
    The standard deviations in the fitted parameters are recorded for both taxonomies.
    The mean of these standard deviations is calculated across the two taxonomies, weighted by the taxonomic proportion of the NEA population, as the final aspect uncertainties.
    These uncertainties should then be added in quadrature with any other uncertainties in the phase curve (e.g. statistical fit uncertainties).
    The latest release of the code can be accessed and downloaded from \href{https://doi.org/10.5281/zenodo.6062631}{doi:10.5281/zenodo.6062631}, where some example input files are provided alongside instructions on compiling and running the code.
    For objects where minor constraints on the shape are known (i.e. a minimum light curve amplitude giving a minimum $a/b$ axis ratio) then this minimum light curve amplitude can be provided to the code to only generate shapes that match this criteria.
    For objects with a well-constrained shape and pole, the methods outlined by \citep{2020A&A...642A.138M} to obtain the `ideal phase curve' are more suitable.
    If the spectral type of the asteroid is known, the code can be edited to provide suitable Hapke parameters for the simulations.
    
    As an example of the significance of this additional uncertainty, we present updated uncertainties for the two example phase curves from \citet{2021PASP..133g5003J}.
    In that work, the authors report the uncertainties for (8014) 1990 MF as $\sigma_H = 0.047$ mag and $\sigma_G = 0.024$.
    Using the process outlined in this work, we derive aspect-related uncertainties of $\sigma_{H\text{, aspect}} = 0.15$ mag and $\sigma_{G\text{, aspect}} = 0.07$, bringing the total uncertainty in the phase curve parameters to $\sigma_{H\text{, total}} = 0.16$ mag and $\sigma_{G\text{, total}} = 0.07$.
    The uncertainties for (19764) 2000 NF5 are reported as $\sigma_H = 0.047$ mag and $\sigma_G = 0.022$.
    Using the process outlined in this work, we derive aspect-related uncertainties of $\sigma_{H\text{, aspect}} = 0.27$ mag and $\sigma_{G\text{, aspect}} = 0.14$, bringing the total uncertainty in the phase curve parameters to $\sigma_{H\text{, total}} = 0.27$ mag and $\sigma_{G\text{, total}} = 0.14$.
    The aspect-related uncertainties dominate over the statistical fit uncertainties, and therefore should not be neglected when analysing phase curves of individual near-Earth asteroids.

\section{Consequences of Aspect-Induced Variability}\label{sec:consequences}

    As indicated in Section \ref{subsec:unrealisticGeoms}, taxonomic classification of near-Earth asteroids using phase curve data is limited by potential aspect effects.
    Aspect effects in phase curves may also propagate significant extra uncertainties into various other studies of these objects.
    A primary effect will be on brightness predictions for future apparitions, potentially hindering observation planning.
    The effect on brightness predictions could also impact space-based intercept and rendezvous missions which rely on camera imaging for guidance and navigation control.
    Geometric effects have already been shown to affect unresolved reflectance spectroscopy of asteroids \citep{2022Icar..37614806P}.
    
    For most near-Earth asteroids, low phase angle ranges are not commonly well-sampled in the phase curves, making estimations of absolute magnitudes highly uncertain.
    This is compounded by the fact that the slope being fit may be modulated by aspect related changes, introducing much larger uncertainty in the absolute magnitudes derived from NEA phase curves.
    This will have a knock-on effect on the estimation of diameters, as calculated from the absolute magnitudes and geometric albedos according to Equation~\ref{eq:diam} \citep{irasChap4}.
    \begin{equation}\label{eq:diam}
        D_\text{eq.} = \frac{10^{-H/5}\cdot1329}{\sqrt{p_{V}}}.
    \end{equation}
    Accurate determination of diameters is important for a range of studies of individual objects.
    Collisional dynamics, both in the outer solar system and for planetary defense purposes, will depend heavily on the size of these objects.
    To calculate the total delta-V required for deflection missions of potentially hazardous asteroids, accurate diameters are needed in conjunction with typical bulk densities to estimate the masses of the objects.
    The ability of the Probabilistic Asteroid Impact Risk (PAIR) model to estimate impact consequences also relies on input distributions for $H$ \citep{2017Icar..289..106M}, and will therefore be limited further by the increased uncertainty in $H$ for a given impactor if only photometric data are available.
    Yarkovsky and YORP modelling of individual objects both depend on estimated diameters as $\propto \frac{1}{D}$ and $\propto\frac{1}{D^2}$ respectively \citep{2015aste.book..509V}.
    Consequently both of these have strong dependence on absolute magnitudes for objects where only photometric data are available.
    Once again this is important in the realm of planetary defense, as if we take the example of (99942) Apophis then predictions of the Yarkovsky drift of this object are needed to constrain the evolution of the orbit beyond the close encounter in 2029 \citep{2013Icar..224..192F}.
    This object has a radar derived diameter, although for a less-studied object the uncertainty in the diameter may begin to dominate in the Yarkovsky drift modelling.
    Phase curves can also be used to detect activity from `active asteroids' \citep[e.g.,][]{2020PSJ.....1...10M}.
    Systematic errors in predicted brightness of these objects can therefore significantly over or under-estimate the level of activity on these objects.
    
    Another area where aspect effects may disrupt accurate modelling is in the infrared regime.
    Thermophysical models such as the NEATM \citep{1998Icar..131..291H} use phase curve parameters in the modelling process to extract 
    physical properties of asteroids.
    From the thermal modelling process using NEATM outlined in \citet{2019A&A...627A.172R}, the geometric albedo ($p_V$) is re-arranged in terms of the effective diameter and absolute magnitude,
    and the Bond albedo ($A_B$) is written in terms of the phase curve slope parameter and the geometric albedo.
    The sub-solar temperature ($T_\text{SS}$) in the NEATM can therefore be re-arranged as follows:
    \small\begin{equation}
        T_\text{SS} = \left[ \frac{\left(1 - (0.290 + 0.684G)\cdot \left( \frac{10^{-H/5}\cdot1329}{D_\text{eq.}} \right)^2\right) \cdot S_{\odot}}{\eta\epsilon\sigma} \right]^{1/4},
    \end{equation}\normalsize
    where $S_\odot$ is the integrated solar flux at the distance of the asteroid, $\epsilon$ is the emissivity, $\eta$ is the thermal beaming parameter, $\sigma$ is the Stefan-Boltzmann constant, $D_\text{eq.}$ is the diameter of an equivalent volume spherical asteroid, and $H$ \& $G$ are the phase curve parameters as usual.
    The output distributions of $p_V$ and $A_B$ will heavily depend on the input distributions for $H$ and $G$ in the bootstrap process, and any increase in the uncertainty in the input distributions will propagate through and increase the uncertainty in  thermally-derived geometric albedos and beaming parameters from the NEATM.
    
    As an example of the consequences of this additional uncertainty on geometric albedos derived from NEATM, we perform thermal modelling of (19764) 2000 NF5 using WISE data taken in the W3 and W4 bands.
    For objects where phase curve parameters were not known, the WISE team assume values and uncertainties according to the first line of Table~\ref{tab:19764NEATM}.
    Doing so results in a thermally derived geometric albedo, $p_V = 0.270 \pm 0.080$.
    \citet{2021PASP..133g5003J} derived a phase curve for this asteroid, and so the input distributions of $H$ and $G$ were defined using the measured parameters as the means, and the statistical fit uncertainties used as the standard deviations (second line of Table~\ref{tab:19764NEATM}).
    This analysis gives a geometric albedo $p_V = 0.211 \pm 0.021$, within the uncertainty range of the values derived using the WISE-assumed phase curve parameters.
    In theory, this has allowed us to better constrain the geometric albedo, reducing the uncertainty significantly.
    However, this does not account for the aspect related uncertainties in the phase curve over this geometry.
    We therefore re-analyse the thermal data using the combined statistical and aspect uncertainties derived in Section~\ref{sec:uncertainty} (also provided in the third line of Table~\ref{tab:19764NEATM}), giving a value $p_V = 0.217 \pm 0.058$.
    
    The percentage uncertainty in the geometric albedo drops significantly if we use derived phase curve parameters in the fitting.
    However, once we properly account for the systematic uncertainty of aspect effects, the percentage uncertainty becomes comparable to that using the assumed values commonly used by the WISE team.
    We can therefore not say that we have constrained the properties of this asteroid better with a derived phase curve than without any phase curve data.
    Aspect-related uncertainty in the phase curves should therefore be accounted for when performing analysis of thermal IR data to avoid over-fitting and underestimating systematics.
    
    \begin{table}\scriptsize
        \begin{tabular}{ccccccr} \hline
            $H$ & $\sigma_{H}$ & $G$ & $\sigma_{G}$ & $p_{V}$ & $\sigma_{P_{V}}$ & HG \& Unc. References \\
            (mag) & (mag) & & & & & \\\hline\hline
            $15.8$ & $0.3$ & $0.15$ & $0.1$ & $0.271$ & $0.08$ & \citep{2011ApJ...743..156M} \\
            $16.028$ & $0.047$ & $0.094$ & $0.022$ & $0.211$ & $0.021$ & \citep{2021PASP..133g5003J} \\
            $16.028$ & $0.27$ & $0.094$ & $0.14$ & $0.213$ & $0.058$ & This Work \\ \hline
        \end{tabular}
        \caption{Various NEATM fits to WISE W3 and W4 data for NEA (19764) 2000 NF5. Initial fit performed using assumed $H$ and $G$ parameters and uncertainties as used by \citet{2011ApJ...743..156M}, second fit using derived phase curve parameters from \citet{2021PASP..133g5003J}, and the third fit using the same parameters but with aspect uncertainty added in quadrature to the statistical fit uncertainties.}
        \label{tab:19764NEATM}
    \end{table}
    
    The limited predictive power of the phase curves for brightness predictions also has further implications for thermal modelling in the infrared transition region ($2.5 - 5\,\si{\micro\metre}$).
    In this region the IR flux at the detector consists of both reflected and emitted components.
    In order to model physical properties of the surface from the emitted component, the reflected flux must first be removed from the observations.
    This involves using $H$ and $G$ to predict the brightness of the asteroid at the given phase angle, and estimating the reflected flux component as \citep[Eq.~A1,][]{2018MNRAS.477.1782R}
    \begin{equation}
        F_{\text{REF},\lambda} = \frac{p_\lambda}{p_V}S_\lambda 10^{\frac{V_\text{SUN} - V}{2.5}},
    \end{equation}
    where $F_{\text{REF},\lambda}$ is the reflected component of the IR flux, $p_\lambda$ is the geometric albedo of the asteroid at the observed wavelength, $p_V$ is the geometric albedo in the Johnson V-band, $S_\lambda$ is the solar flux at $1$\,au at the observed wavelength, $V_\text{SUN}$ is the Johnson V magnitude of the Sun at 1\,au, and $V$ is the predicted (using the $H$ and $G$ phase curve parameters) apparent magnitude of the asteroid as viewed from Earth.
    \citet{2018MNRAS.477.1782R} find that under some geometries, the reflected component can account for up to several tens of per cent of the total IR flux received at the detector.
    These brightness predictions from the phase curves have the ability to massively over- or under-estimate the brightness of the asteroid at a given geometry, and therefore heavily modify the modelled parameters from the remaining emitted IR flux.
    
    We analyse the near-IR Spitzer observations of (159402) 1999 AP10 in the $3.6\,\si{\micro\metre}$ and $4.5\,\si{\micro\metre}$ bands \citep{2010AJ....140..770T}, and calculate the expected flux contribution of reflected light.
    To derive the reflected flux components in each band, and their corresponding uncertainties, we use: the phase curve parameters and uncertainties used by the Spitzer team \citep{2010AJ....140..770T}, and the parameters derived from our phase curve data with both the statistical phase curve fit uncertainties and the combined (statistical and aspect) uncertainties.
    We also present the estimate of the reflected flux component when using an incorrect value for $G$, as an illustration of the significant effect of changes in this parameter.
    These results are summarised in Table \ref{tab:reflectedFlux}.
    
    \begin{table*}
        \begin{tabular}{ccccccccr} \hline
            $H$ & $\sigma_{H}$ & $G$ & $\sigma_{G}$ & $F_{\text{REF},3.6\si{\micro\metre}}$ & $\sigma_{F_{\text{REF},3.6\si{\micro\metre}}}$ & $F_{\text{REF},4.5\si{\micro\metre}}$ & $\sigma_{F_{\text{REF},4.5\si{\micro\metre}}}$ & HG \& Unc. Reference \\
            (mag) & (mag) & & & (\%) & (\%) & (\%) & (\%) & \\\hline\hline
            $16.443$ & $0.5$   & $0.15$  & $0.0$   & $76.565$  & $36.965$ & $15.172$ & $7.325$ & \citep{2010AJ....140..770T}\\ 
            $16.321$ & $0.025$ & $0.023$ & $0.011$ & $55.785$  & $2.112$  & $11.054$ & $0.418$ & This work\\
            $16.321$ & $0.34$  & $0.023$ & $0.27$  & $61.051$  & $45.614$ & $12.098$ & $9.039$ & This work\\
            $16.321$ & $0.025$ & $0.5$   & $0.011$ & $126.154$ & $3.408$  & $24.998$ & $0.675$ & This work (H only)\\ \hline& 
        \end{tabular}
        \caption{Estimates of the reflected flux components of the detected infrared flux for (159402) 1999 AP10. Evaluated for the \textit{Spitzer} $3.6\si{\micro\metre}$ and $4.5\si{\micro\metre}$ bands within the infrared transition region. First estimate uses the values assumed by the \textit{Spitzer} team for the phase curve parameters. The second estimate uses the phase curve parameters determined from phase curve observations with PIRATE, with statistical fit uncertainties. The third estimate uses the combined aspect and statistical fit uncertainties for the phase curve parameters. The fourth estimate uses a wildly wrong value of G to signify the impact of picking arbitrary values for the phase curve parameters. The choice of phase curve parameters to predict the V magnitude, and subsequently convert to an IR flux is extremely important. Using values from a measured phase curve can better constrain the percentage reflected flux. However, once we include the potential aspect uncertainty in the phase curve parameters it can be seen that this is not as well constrained as initially thought.}
        \label{tab:reflectedFlux}
    \end{table*}
    
    If we were to use the derived shape model and assumed S-type Hapke parameters to estimate the brightness at the geometry over which this asteroid was observed by Spitzer, we obtain $F_{\text{REF},3.6\si{\micro\metre}} = 63$ per cent and $F_{\text{REF},4.5\si{\micro\metre}} = 12$ per cent.
    However, to get a proper prediction with this model we would need to derive accurate Hapke parameters instead of using assumed values based on the taxonomy of the asteroid.
    This is not typically feasible for near-Earth asteroids, meaning phase curve estimates of the brightness are usually needed.
    Using a significantly incorrect value for $G$ can lead to a conclusion that the reflected flux component is larger than the total flux received at the detector, which can not be true.
    We must therefore at least try to make an informed decision on the phase curve parameters to use before doing this kind of modelling, rather than using one set of slope parameters across the entire asteroid population.
    Using the Spitzer team values for the phase curve parameters (using an assumed $G$ value), the estimate of the reflected flux component is significantly higher in both bands than using the parameters derived from our phase curve data.
    Ignoring the systematic effects of aspect changes, we could think that we have constrained this reflected flux component very well.
    However, when we include the aspect uncertainty in our Monte Carlo estimation we see that these values are very poorly constrained.
    Ignoring the additional uncertainty that may arise from aspect changes will therefore significantly underestimate the uncertainty in the thermally-emitted component of the IR flux at the detector, and consequently underestimate the uncertainty in any physical properties derived from thermophysical modelling thereafter.
    
\section{Conclusions}\label{sec:conclusions}
    
    In this paper, we have investigated the effect of aspect changes on the phase curves of near-Earth asteroids.
    We have presented an investigation into the variability in the phase curve of near-Earth asteroid (159402) 1999 AP10.
    The light curves from the PIRATE facility are combined with data from Palmer Divide Observatory obtained from ALCDEF, in order to produce a convex shape model and pole solution.
    The derived shape, pole, and sidereal rotation period match expectations from independent data.
    This model is then used to verify that the deviations in the phase curve arise from aspect effects instead of random scatter in the data.
    This is achieved via simulation of the phase curve over the apparition using the shape model and a Hapke photometric model, which accurately recreates the variation in the phase curve.
    The flux-weighted mean latitude of visible and illuminated facets on the shape model (the `flux latitude') changes from the southern hemisphere to the northern hemisphere throughout the apparition.
    This change in observed hemisphere drives the aspect-related effects in the phase curve due to the difference in shape between the two hemispheres of this asteroid.
    A large change in flux latitude may accompany large changes in light curve morphology, so observed changes in the light curves used to construct a phase curve may indicate the potential presence of aspect changes due to changing flux latitude.
    However, a lack of light curve morphological changes does not necessarily mean a small change in flux latitude or observed cross-section, so care must still be taken to consider this effect.
    
    After the detection of these aspect changes in the phase curve of (159402) 1999 AP10, a theoretical study was undertaken in order to understand how this effect may affect the phase curves of other asteroids.
    Phase curves are simulated for a wide variety of ellipsoidal shape models over different viewing geometries, demonstrating some of the worst-case scenarios that may arise.
    A representative sample of asteroid models is simulated over the 2020 apparition of (19764) 2000 NF5, displaying the strong dependence of aspect related modulations to the phase curve on the rotation pole orientation and change in flux latitude.
    
    Aspect effects introduce a systematic uncertainty in the $H$ and $G$ phase curve parameters fit to the data.
    These additional uncertainties are highly dependent on specific viewing geometry, so care must be taken to model the potential impact of this effect for each object when using parameters derived from their phase curves in further studies.
    Taxonomic classification using phase curve data is deemed unreliable from phase curves without supplemental data to support the classification.
    We provide a program for estimating the additional uncertainty introduced into estimations of $H$ and $G$ when the shape and pole of the object is unknown.
    
    We have discussed the effect of this additional aspect uncertainty on wider studies of individual asteroids where convex inversion and extraction of the surface scattering properties directly is not feasible.
    This can propagate through to estimations of the diameter, affecting Yarkovsky and YORP effect modelling.
    We have shown how the additional aspect uncertainty in the phase curve parameters can affect derivations of asteroid properties such as the geometric albedo, and also how this affects our ability to properly separate reflected and emitted fluxes in observations of asteroids in the infrared transition region.
    
    This study has considered this effect over a wide variety of convex shapes.
    However, real asteroid shapes may have significant concavities and the aspect effects may be over or underestimated still by these methods.
    The information from non-convex shapes is typically lost when rotationally averaging for phase curves, however.
    It remains the subject of future work to determine the significance that large concavities may have on this effect.

\section*{Acknowledgements}

    S.~L. Jackson is funded by the Science and Technology Facilities Council under grant ST/T506321/1 (project reference: 2284918).
    B. Rozitis is funded by the Science and Technology Facilities Council under grant ST/T000228/1.
    L.~R. Dover is funded by the Science and Technology Facilities Council under grant ST/S505456/1 (project reference: 2156534).
    S.~F. Green is funded by the Science and Technology Facilities Council under grant ST/T000228/1
    U.~C. Kolb is funded by the Science and Technology Facilities Council under grant ST/T000295/1.
    A.~E. Andrews is funded by the Science and Technology Facilities Council under grant ST/R504993/1 (project reference: 1946865).
    S.~C. Lowry is funded by the Science and Technology Facilities Council under grant ST/S000348/1.
    
    The authors thank the reviewer, Dr. Tomasz Kwiatkowski, for their constructive and helpful comments on an earlier version of this manuscript.
    This research was made possible through the OpenSTEM Labs, an initiative funded by HEFCE and by the Wolfson Foundation.
    The authors thank Sybilla Technologies, Baader Planetarium, and the Instituto de Astrof\'isica de Canarias (IAC) for their assistance developing new capabilities and supporting the continuing operation of the OpenScience Observatories.
    This work uses data obtained from the Asteroid Lightcurve Data Exchange Format (ALCDEF) database, which is supported by funding from NASA grant 80NSSC18K0851.

\section*{Data Availability}
 
The PIRATE light curves of (159402) 1999 AP10 will be made available on Asteroid Lightcurve Data Exchange Format (ALCDEF) database\footnote{\url{https://minplanobs.org/alcdef/}} \citep{2011MPBu...38..172W}.
The generated shape model of (159402) 1999 AP10 will be made available through the Database of Asteroid Models from Inversion Techniques\footnote{Available at: \url{https://astro.troja.mff.cuni.cz/projects/damit/}} \citep[DAMIT;][]{2010A&A...513A..46D}.
Aside from the previously mentioned repositories; the light curve data, phase curve data, and the shape model for (159402) 1999 AP10 are available on Open Research Data Online\footnote{\url{https://ordo.open.ac.uk/collections/Near-Earth_Asteroid_159402_1999_AP10_Light_Curve_Data_Phase_Curve_Data_and_Shape_Model/5868821/1}} \citep{AP10_Data}.
The latest release of the aspect code is available at \href{https://doi.org/10.5281/zenodo.6062631}{10.5281/zenodo.6062631}.



\bibliographystyle{mnras}
\bibliography{references} 


\appendix

\section{Mathematical Formulation of Hapke Photometric Model}\label{app:hapke}

    To define the components of the bidirectional reflectance function, we first define some additional helper equations as follows to simplify equations found later in this section.
    Equations defined herein are reproduced from \citet{hapke2012theory}.
    \begin{align}
        &f(\psi) = \exp\left( -2\tan{\frac{\psi}{2}} \right), \\
        &\chi(\bar{\theta}) = \left( 1 + \pi\tan^2{\bar{\theta}} \right), \\
        &E_1(y) = \exp\left( - \frac{2}{\pi}\cot{\bar{\theta}}\cot{y} \right), \\
        &E_2(y) = \exp\left( - \frac{1}{\pi}\cot^2{\bar{\theta}}\cot^2{y} \right), \\
        &\eta(y) = \chi(\bar{\theta})\left[ \cos{y} + \sin{y}\tan{\bar{\theta}}\frac{E_2(y)}{2 - E_1(y)} \right]. 
    \end{align}
    where $y$ can be either $i$ or $e$.
    It should be noted that the above equations are only defined to simplify other equations in this section, and do not carry any significant physical meaning on their own.
    In the case where $i \leq e$, the effective cosines and the rough surface correction to the bidirectional reflectance are defined as
    \begin{align}
        &\mu_{0e} = \chi(\bar{\theta}) \left[ \mu_0 + \sin{i}\tan{\bar{\theta}}\frac{\cos{\psi}E_2(e) - \sin^2\left( \frac{\psi}{2} \right)E_2(i)}{2 - E_1(e) - \left( \frac{\psi}{\pi} \right)E_1(i)} \right], \\
        &\mu_e = \chi(\bar{\theta}) \left[ \mu + \sin{e}\tan{\bar{\theta}}\frac{E_2(e) - \sin^2\left( \frac{\psi}{2} \right)E_2(i)}{2 - E_1(e) - \left( \frac{\psi}{\pi} \right)E_1(i)} \right], \\
        &S(i,e,\psi,\bar{\omega}) = \frac{\mu_e}{\eta(e)}\frac{\mu_0}{\eta(i)}\frac{\chi(\bar{\theta})}{1 - f(\psi) + f(\psi)\chi(\bar{\theta})\frac{\mu_0}{\eta(i)}}. 
    \end{align}
    In the case where $i > e$, these parameters are defined as
    \begin{align}
        &\mu_{0e} = \chi(\bar{\theta}) \left[ \mu_0 + \sin{i}\tan{\bar{\theta}}\frac{E_2(i) - \sin^2\left( \frac{\psi}{2} \right)E_2(e)}{2 - E_1(i) - \left( \frac{\psi}{\pi} \right)E_1(e)} \right], \\
        &\mu_e = \chi(\bar{\theta}) \left[ \mu + \sin{e}\tan{\bar{\theta}}\frac{\cos{\psi}E_2(i) - \sin^2\left( \frac{\psi}{2} \right)E_2(e)}{2 - E_1(i) - \left( \frac{\psi}{\pi} \right)E_1(e)} \right], \\
        &S(i,e,\psi,\bar{\omega}) = \frac{\mu_e}{\eta(e)}\frac{\mu_0}{\eta(i)}\frac{\chi(\bar{\theta})}{1 - f(\psi) + f(\psi)\chi(\bar{\theta})\frac{\mu_0}{\eta(e)}}. 
    \end{align}    
    The opposition effect function is defined as
    \begin{equation}
        B(\alpha,B_0,h) = \frac{B_0}{\left[ 1 + \frac{\tan{\frac{\alpha}{2}}}{h} \right]},
    \end{equation}
    where $h$ corresponds to the width of the opposition surge, and $B_0$ is the amplitude of the opposition effect at a phase angle $\alpha = 0\si\degree$.
    The single-particle phase function is calculated as
    \begin{equation}
        p(\alpha,g) = \frac{1-g^2}{\left( 1 + 2g\cos{\alpha} + g^2 \right)^\frac{3}{2}},
    \end{equation}
    where $g$ is the scattering asymmetry parameter.
    For the asymmetry parameter; $g < 0$ corresponds to back-scattering particles, $g = 0$ corresponds to isotropic scattering, and $g > 0$ corresponds to forward-scattering particles.
    The approximate form of the \citet{1960ratr.book.....C} H-function is calculated as
    \begin{equation}
        H(x,\omega) = \frac{1 + 2x}{1 + 2x\sqrt{1 - \omega}},
    \end{equation}
    where $x$ can be either $\mu_{0e}$ or $\mu_e$.

\bsp	
\label{lastpage}
\end{document}